\documentclass[aps,prl,amsmath,amssymb,twocolumn,showpacs,showkeys,superscriptaddress, floatfix]{revtex4-1}

\usepackage{graphicx}
\usepackage{dcolumn}
\usepackage{color}
\usepackage{ulem}
\usepackage{float}
\usepackage{hyperref}
\usepackage{nameref}

\begin{document}

\title{Modification of optical response by subsequent growth of In(Ga)As(Sb) structures on GaP substrate}


\title{Optical response of (InGa)(AsSb)/GaAs quantum dots embedded in a GaP matrix}

\author{Petr Steindl}
\email[]{steindl@physics.leidenuniv.nl}
\affiliation{Department of Condensed Matter Physics, Faculty of Science, Masaryk University, Kotl\'a\v{r}sk\'a~267/2, 61137~Brno, Czech~Republic}
\affiliation{Central European Institute of Technology, Masaryk University, Kamenice 753/5, 62500~Brno, Czech~Republic}
\affiliation{Huygens-Kamerlingh Onnes Laboratory, Leiden University, P.O. Box 9504, 2300 RA Leiden, Netherlands}

\author{Elisa Maddalena Sala}
\email[]{e.m.sala@sheffield.ac.uk}
\affiliation{Center for Nanophotonics, Institute for Solid State Physics, Technische Universit\"{a}t Berlin, Germany}
\affiliation{EPSRC National Epitaxy Facility, The University of Sheffield, North Campus, Broad Lane, S3 7HQ Sheffield, United Kingdom}

\author{Benito Al\'{e}n}
\affiliation{Instituto de Micro y Nanotecnolog\'{i}a, IMN-CNM, CSIC (CEI UAM+CSIC) Isaac Newton, 8, E-28760, Tres Cantos, Madrid, Spain}

\author{David Fuertes Marr\'{o}n}
\affiliation{Instituto de Energ\'{i}a Solar (IES), Universidad Polit\'{e}cnica de Madrid, Avda. Complutense 30, 28040 Madrid, Spain}


\author{Dieter Bimberg}
\affiliation{Center for Nanophotonics, Institute for Solid State Physics, Technische Universit\"{a}t Berlin, Germany}
\affiliation{"Bimberg Chinese-German Center for Green Photonics'' of the Chinese Academy of Sciences at CIOMP, 13033 Changchun, China}
%

\author{Petr Klenovsk\'y}
\email[]{klenovsky@physics.muni.cz}
\affiliation{Department of Condensed Matter Physics, Faculty of Science, Masaryk University, Kotl\'a\v{r}sk\'a~267/2, 61137~Brno, Czech~Republic}
\affiliation{Central European Institute of Technology, Masaryk University, Kamenice 753/5, 62500~Brno, Czech~Republic}
\affiliation{Czech Metrology Institute, Okru\v{z}n\'i 31, 63800~Brno, Czech~Republic}

\date{\today}

\begin{abstract}
The optical response of (InGa)(AsSb)/GaAs quantum dots (QDs) grown on GaP (001) substrates is studied by means of excitation and temperature-dependent photoluminescence (PL), and it is related to their complex electronic structure. Such QDs exhibit concurrently direct and indirect transitions, which allows the swapping of $\Gamma$ and $L$ quantum confined states in energy, depending on details of their stoichiometry. Based on realistic data on QD structure and composition, derived from high-resolution transmission electron microscopy (HRTEM) measurements, simulations by means of $\mathbf{k\cdot p}$ theory are performed. The theoretical prediction of both momentum direct and indirect type-I optical transitions are confirmed by the experiments presented here. Additional investigations by a combination of Raman and photoreflectance spectroscopy show modifications of the hydrostatic strain in the QD layer, depending on the sequential addition of QDs and capping layer. A variation of the excitation density across four orders of magnitude reveals a 50\,meV energy blueshift of the QD emission. Our findings suggest that the assignment of the type of transition, based solely by the observation of a blueshift with increased pumping, is insufficient. We propose therefore a more consistent approach based on the analysis of the character of the blueshift evolution with optical pumping, which employs a numerical model based on a semi-self-consistent configuration interaction method.
\end{abstract}

%
\vspace{0.5cm}

\maketitle



\tableofcontents

\setcounter{secnumdepth}{3}


\section{Introduction}

The growth and the physical properties of III-V quantum dots (QDs) have been extensively studied, leading to a variety of appealing applications, especially in semiconductor opto-electronics. Such QDs are crucial for classical telecommunication devices as for low threshold/high bandwidth semiconductor lasers and amplifiers~\citep{Bimberg1997,Ledentsov,Heinrichsdorff1997,Schmeckebier2017, Unrau_laserphotonics_2014}, and for single photon and entangled photon pair emitters for quantum communication~\cite{yuan_electrically_2002, martin-sanchez_single_2009, salter_entangled-light-emitting_2010,takemoto_quantum_2015, schlehahn_single-photon_2015, kim_two-photon_2016, paul_single-photon_2017, muller_quantum_2018, Plumhof2012,Trotta:16, Aberl_PRB2017, KLenovsky_PRB2018}, among other quantum information technologies \cite{li_all-optical_2003, robledo_conditional_2008, kim_quantum_2013, yamamoto_present_2011, michler_quantum_2017, Krapek2010, Klenovsky2016, Kindel_prb2010}. Most of the present applications in optics are based on so-called type-I QDs, which show direct electron-hole recombination in both real and ${\bf k}$-space, as for In(Ga)As QDs embedded in a GaAs matrix. 
Much less attention has been given to type-I indirect and/or type-II QDs, particularly antimony-based ones, like In(Ga)As QDs overgrown by a thin Ga(AsSb) layer~\cite{ripalda_room_2005, liu_room-temperature_2006, ulloa_high_2012, Klenovsky2010, Klenovsky_IOP2010, Klenovsky2015}, or In(Ga)Sb QDs in a GaAs matrix~\cite{hayne_electron_2003, alonso_optical_2007, tatebayashi_lasing_2007, laghumavarapu_gasbgaas_2007, Hodgson_SST2018_InGaSbQDs}, which show spatially indirect optical transitions. 
Such structures generally require more challenging growth processes, but bring new and improved characteristics, for example intense room temperature emission~\cite{ripalda_enhancement_2007}, naturally low fine-structure splitting (FSS)~\cite{Krapek2015}, increased tuneability of the exciton confinement geometry and topology~\cite{young_optical_2012,llorens_type_2015,Klenovsky2016,llorens_wave-function_2018}, radiative lifetime~\cite{Gradkowski2012, Hodgson_jap2016} and magnetic properties~\cite{llorens_topology_2019, Hayne_apl2003,Bansal_prb2008}. 

The use of GaP as matrix material for III-V QDs has recently attracted particular attention due to the possibility of defect-free growth on silicon since the lattice mismatch between GaP and Si amounts only to 0.4\,\%~\cite{Grassman_apl2013}. Thus, the integration of III-V QD-based opto-electronic devices with silicon-based ones is feasible~\cite{Beyer_jap2013}. Moreover, (InGa)(AsSb)/GaP QDs, due to their huge hole localization energies, result in long hole storage times and can be utilized as building blocks for a novel kind of nanoscale memory, the QD-Flash~\cite{Sala2018, Marent2011, Hodgson_jap2013, Hayne_2013, Young_apl2012}. However, the growth of defect-free systems, in particular by the most important mass production process MOCVD (Metal Organic Chemical Vapour Deposition), is very challenging due to the large lattice mismatch between Sb- and P-based structures (GaAs/GaP 3.6\,\%, $\mathrm{ In_{0.5}Ga_{0.5}As/GaP}$ 7.4\,\%, InAs/GaP 11.5\,\%, GaSb/GaP 11.8\,\%, InSb/GaP 18.9\,\%~\cite{Vurgaftman2001}). Using specific growth engineering of MOCVD, $\mathrm{ In_{0.5}Ga_{0.5}As/GaP}$ QDs have been obtained by Stracke \textit{et al.}~\cite{Stracke_APL2012_GaAsInterlayer, Stracke2014} and a hole storage time at room temperature of about 4 min was reported~\cite{Bonato_APL2015}. Further improvement in the storage time beyond the magic 10 y limit might be obtained for type-II antimony-based QDs in an (AlGa)P matrix~\cite{Bimberg2011_SbQDFlash, Marent2009_microelectronics}. ``10 year limit'' is the minimal retention time of any commercial Flash memory. The novel ``QD-Flash'' memory concept based on type II QDs, combining the best properties of a DRAM with a Flash, which was introduced by one of the co-authors~\cite{Marent_APL2007_10y} has to reach one final milestone, which is a 10 y retention time of the information. 

Recently, localization energy up to 1.15 eV, corresponding to localization time of 1 hr at room temperature, has been obtained for (InGa)(AsSb)/GaAs/AlP/GaP QDs by Sala \textit{et al.}~\cite{t_sala, Sala2018}, which represents to date the record for MOCVD-grown QDs. Until now, the longest published storage time is 4~days for holes trapped in GaSb/GaP QDs grown by Molecular Beam Epitaxy (MBE)~\cite{Bonato2016_PSSB}. However, long growth times and associated high costs of MBE might render this growth approach prohibitive for large scale industrial use, thus favoring MOCVD growth.

\begin{figure}[h]
\centering
\includegraphics[width=\linewidth]{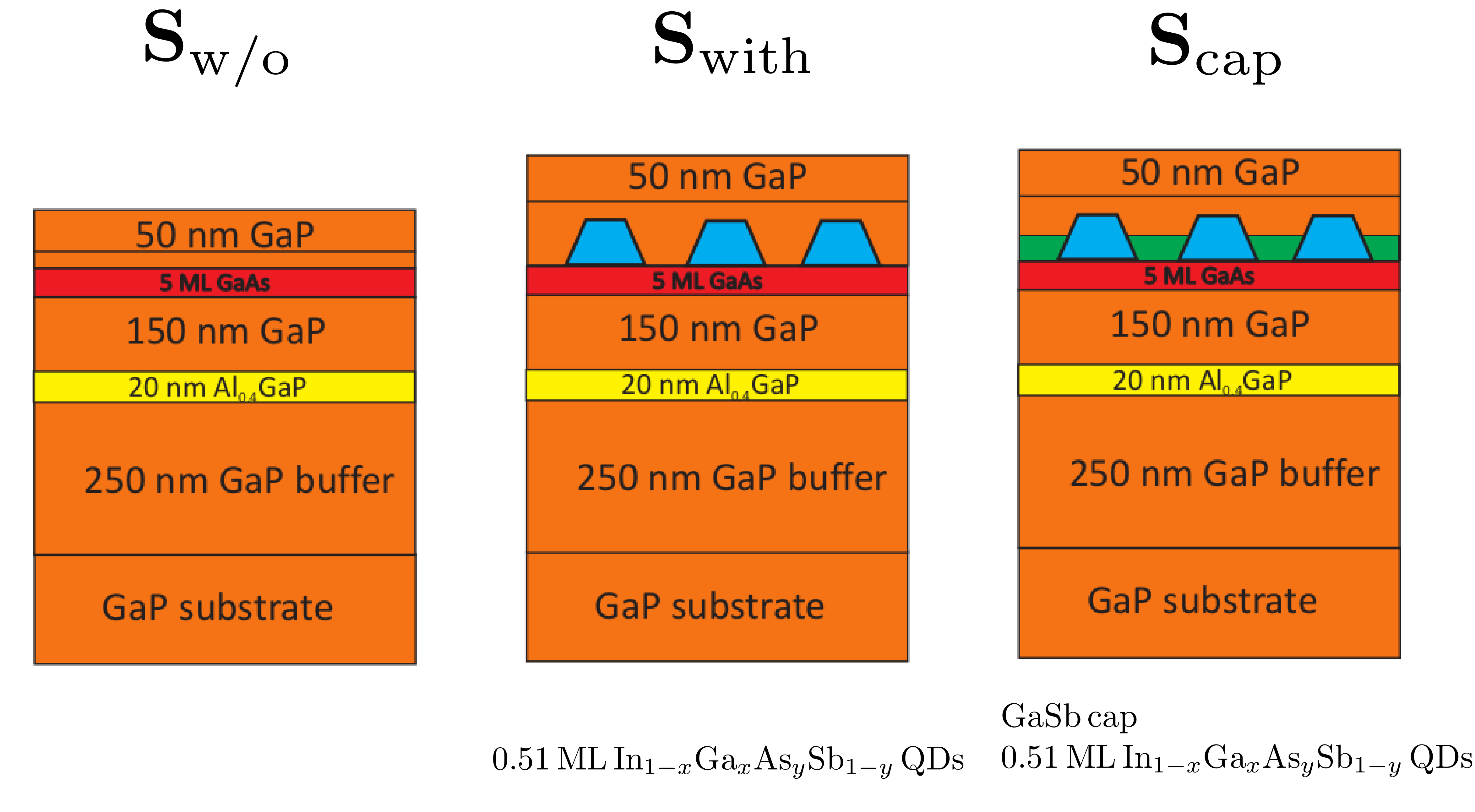}
\caption{Set of samples studied in this work: $S_\mathrm{w/o}$ represents the structure without QDs (only GaAs interlayer), $S_\mathrm{with}$ with In$_{1-x}$Ga$_{x}$As$_y$Sb$_{1-y}$ QDs on top of the GaAs layer, and $S_\mathrm{cap}$ with an additional GaSb capping layer above the QDs.}
\label{fig:TUstructure}
\end{figure}

\setcounter{tocdepth}{1}

In this work we study the optical transitions of antimony-based III--V ~In$_{1-x}$Ga$_x$As$_y$Sb$_{1-y}$/GaAs QDs embedded in a GaP matrix by means of excitation and temperature resolved photoluminescence (PL). The experimental results are compared to simulations based on $\mathbf{k\cdot p}$ theory, enabling us to distinguish more easily direct and indirect optical transitions. The manuscript is organized as follows: in the next section, the growth details and structural characterization results of our nanostructures are presented. High-resolution transmission electron microscopy (HRTEM) provides an insight into the QD structure and material distribution. Next, Raman measurements allow to estimate the strain in the QD areas. The effect of strain on the {\bf k}-direct transitions is studied by photoreflectance measurements.
$\mathrm{{\bf k}}\cdot{ \mathrm{\bf p}}$ calculations are then presented, based on the real shape of QDs and composition variations. Eventually, intensity and temperature dependent PL spectra of individual samples are studied. Finally, the polarization anisotropy of emission from different samples is discussed.
%

\section{Sample fabrication and structural characterization} \label{sec:Fabrication}
\begin{figure*}[!ht]
	\centering
	\includegraphics[width=1\linewidth]{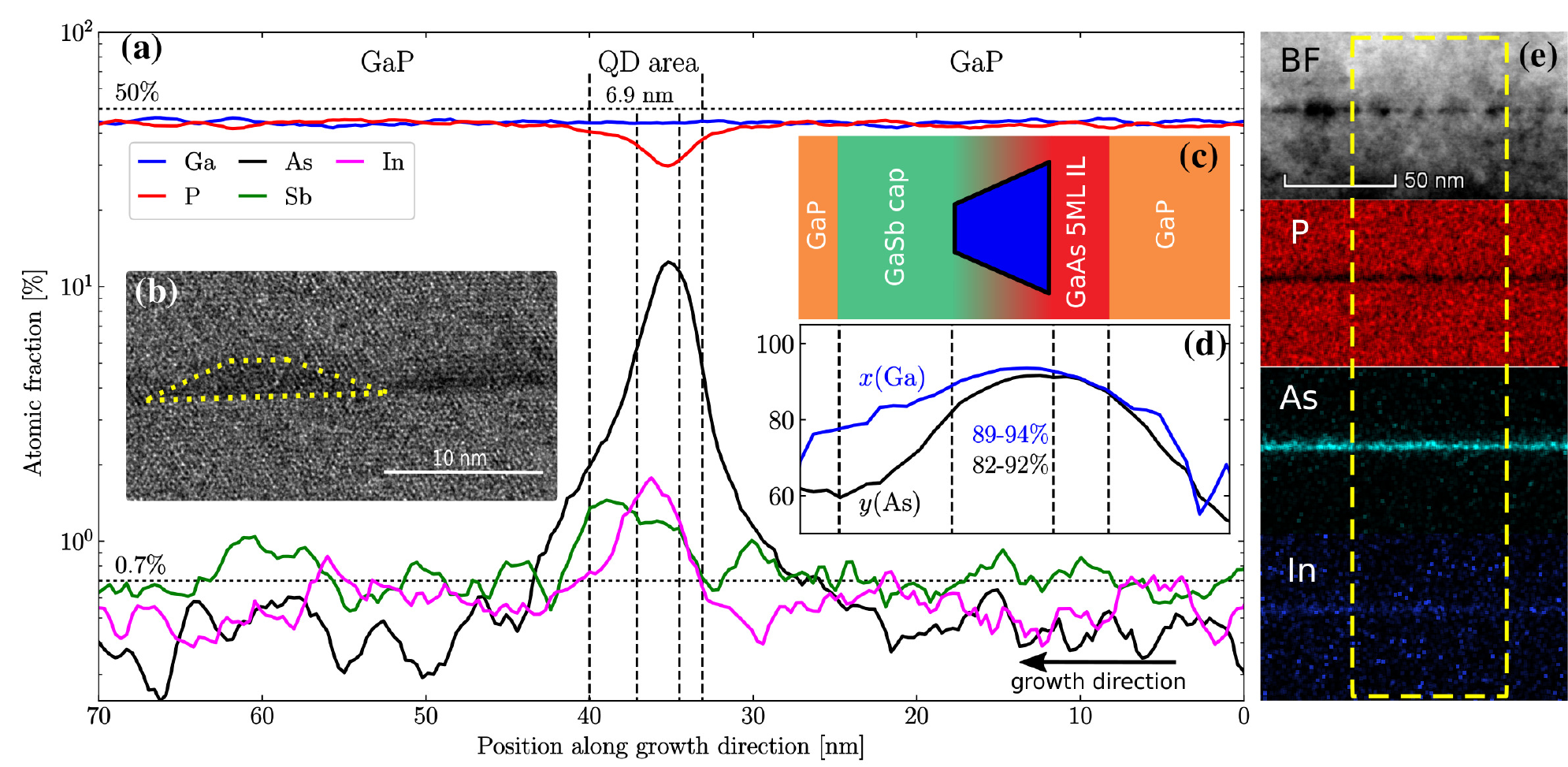}
	\caption{
	{$\bf{(a)}$} Material distribution profile taken by EDX along the growth direction for all elements present in sample $S_\mathrm{cap}$, and averaged over the area highlighted by the yellow dashed curve of panel {$\bf{(d)}$}. In the QD area, between the vertical positions of 30 and 40$\,$nm in the cut, a substantial increase of In, As, Sb and a reduction of P is observed. {$\bf{(b)}$} HRTEM micrograph clearly showing a single QD with truncated-pyramid shape (highlighted in yellow). The inset {$\bf{(c)}$} shows a schematic of the investigated sample, while inset {$\bf{(d)}$} shows the estimated composition of Ga and As around the QD depicted in {$\bf{(c)}$}. The panel {$\bf{(e)}$} displays, from top to bottom: cross-section HRTEM image of $S_\mathrm{cap}$ and three EDX concentration profiles measured at the same time: red for P, white for As and blue for In. HRTEM images were taken under strong-beam bright field condition using the (200) reflection perpendicular to the growth direction.}
	\label{fig:TEM}
\end{figure*}

A schematic depiction of the samples studied in this work is presented in Fig.~\ref{fig:TUstructure}. They have been grown by MOCVD on GaP(001) substrates, in a horizontal Aixton 200 reactor, using H$_2$ as carrier gas. The growth of~the~In$_{1-x}$Ga$_x$As$_y$Sb$_{1-y}$ QDs is based on the Stranski-Krastanov mode~\cite{Wang_2002_ExchangeAnion} and requires a few-ML-thick GaAs interlayer, which will be denoted here as IL. The growth of such material system has been previously studied by Sala \textit{et al.} in~\cite{Sala2018,t_sala, Sala2016}. The growth procedure starts with a 250$\,$nm GaP buffer layer, followed by a 20$\,$nm Al$_{0.4}$Ga$_{0.6}$P layer providing a barrier for the photogenerated charge carriers, and 150$\,$nm GaP at a temperature of 750$\,^\circ$ C. The substrate temperature is then reduced to 500$\,^\circ\rm{C}$ and the following steps are carried out: $\textit{(i)}$ growth of a 5$\,$ML-thick GaAs interlayer, required for QD formation ~\cite{Sala2018, t_sala}, $\textit{(ii)}$  a short Sb-flush by supplying Triethylantimony for the QD samples S$_\mathrm{with}$ and S$_\mathrm{cap}$, with a flux of $2.6\,\mathrm{\mu mol/min}$, $\textit{(iii)}$ nominally $ \sim$ 0.51~ML $\mathrm{ In_{0.5}Ga_{0.5}Sb}$ QDs, $\textit{(iv)}$ a 1$\,$ML thick GaSb cap for the sample S$_\mathrm{cap}$, $\textit{(v)}$ a growth interruptions (GRI) of 1$\,$s without any precursor supply, and $\textit{(vi)}$ an additional GaP cap layer $\sim$ 6$\,$nm thick (thickness optimized to maximize PL intensity of the structure, see~ Sect.~5.8~of~Ref.~\citep{t_sala}). Finally, the samples are heated again to 620$\,^\circ\rm{C}$ for the growth of a 50$\,$nm GaP spacer.

We would like to point out that the optimum growth temperature for the GaP spacer was previously investigated to suppress thermally activated In and/or Ga interdiffusion~\citep{t_sala}. For instance, a temperature equal or greater than 650$\,^{\circ}$C can lead to blueshift of the QD emission, as reported for InAs/GaAs QDs~\cite{Xu_APL1998}, \cite{Malik_APL1997}: at high temperature In and/or Ga may diffuse from the QD layer across the QDs/matrix interface, leading to changes in QD size and composition, and therefore to a blueshif of the QD emission. Growing a spacer layer after a proper QD capping affects neither the chemical composition of the QDs nor leads to any material intermixing in the QDs. On the other hand, we note that growing such a layer at high temperatures straight after the growth of QDs, could lead to remarkable As/P intermixing, as observed for example for GaAs self-assembled QDs (SADQs) on GaP in~\cite{Abramkin_JAP2012}.

The differences between the three types of samples are summarized in Tab.~\ref{tab:samples}.

\begin{table}[h]
	\caption{Labels of studied samples and their structural differences.}
	\begin{tabular}{cl}
		\hline
		label& \multicolumn{1}{c}{specification}\\ 		
		\hline
		\hline
		 S$_\mathrm{w/o}$ & 5$\,$ML GaAs\\
		S$_\mathrm{with}$ & 5$\,$ML GaAs, 0.51$\,$ML QDs\\
		S$_\mathrm{cap}$ & 5$\,$ML GaAs, 0.51$\,$ML QDs, GaSb cap\\
		\hline
	\end{tabular}\label{tab:samples}
\end{table}

The structural characterization has been carried out by a HRTEM Titan Themis with the add-on SUPER-X energy-dispersive X-ray (EDX) detector. Fig.~\ref{fig:TEM} shows a cross-sectional HRTEM micrograph of sample $S_\mathrm{cap}$, where an additional GaSb cap has been used above the QDs. The EDX analysis has been employed to determine the distribution profile of the constituents in the QDs and the surrounding matrix. The range displayed in Fig.~\ref{fig:TEM} is $\sim$ 70\,nm, centered around the QD region, which comprises the GaAs IL, the QDs, the GaSb layer, and the GaP cap. 

Starting from the right side of Fig.~\ref{fig:TEM}{$\bf{(a)}$}, the concentration profile follows the growth sequence depicted in panel~{$\bf{(c)}$}. Due to the presence of the GaAs IL, an increase of As and a reduction of P is observed. The thickness of the highlighted QD area comprising GaAs IL, QDs and GaSb cap is $\sim$6.9 nm. Solely 5 ML GaAs IL amounts to~1.4\,nm (1\,ML GaAs $\sim$0.27\,nm). The size and shape of the QDs are determined by the HRTEM investigations, and a micrograph of a single QD is depicted in the inset~{$\bf{(b)}$}. The QD shows a truncated-pyramid shape with a base length of about 15$\,$nm and a height of 2.5\,nm. These dimensions are in good agreement with those previously reported by Sala \textit{et al.}~\cite{Sala2018,t_sala}.

QDs discussed here are grown at the low temperature of 500$\,^{\circ}$C. We thus expect to observe very little or no As-P intermixing in the QDs. In fact, as also discussed in~\cite{Shamirzaev_APL2010} for GaAs/GaP SAQDs structures, a reduced growth temperature of 550$\,^{\circ}$C leads to suppression of As-P intermixing. Also, the QDs studied in this work are isolated from the GaP matrix by the GaAs IL, which additionally prevents As-P exchange to take place in the QDs. Note, that this scenario is profoundly different to that observed,~e.g.,~for GaSb QDs grown on bare GaP in \cite{Desplanque_2017} and \cite{Abramkin2014}. There, a considerable Sb-P intermixing during QD formation was observed, which led to a reduction of the lattice mismatch between QDs and matrix, thus enabling the QD growth and preventing the introduction of misfit dislocations. These observations, together with the fact that the GaP cap layer is grown at low temperature, leads us to the conclusion, that we can largely exclude an As-P intermixing in our QDs. However, a slight As-P intermixing at the GaP/GaAs IL interface may be deduced from the P and As profiles, since both As and P are detected in the GaAs IL area.
Nevertheless, it is important to note that the EDX results represent an average over a large area, as already stated before, and cannot be compared with much more precise XSTM measurements, which would provide a more accurate insight on the QD composition and material distribution. Therefore, we can only partially infer the chemical composition of the QD area. 

The maximum concentration of As can be found roughly at the GaAs/In$_{1-x}$Ga$_{x}$As$_y$Sb$_{1-y}$ interface, considering that the IL thickness is 1.4\,nm. As a consequence, it is very likely that a considerable amount of As can be found in QDs, as also previously suggested by Sala \textit{et al.}~\cite{Sala2018, t_sala}. Note that we consider a region where a clear In concentration is detectable (i.e. larger than 0.7$\,$\%, see panel {$\bf{(a)}$}) to be the QD area. Similarly, this has been already observed in XSTM studies on In$_{0.5}$Ga$_{0.5}$As/GaP QDs, where the In concentration was largest inside the QD~\cite{Prohl2013}.

The Sb concentration increases in the segment corresponding to the GaSb cap, while at the same time both In and As concentrations decrease. Simultaneously, the P content slightly increases, possibly due to a slight Sb-P intermixing, since P tends to replace Sb already at low growth temperature, thus creating Ga-P bonds, as also observed in~\cite{Abramkin2014}. It is worth to point out that such increase corresponds to the decrease of the In peak concentration, i.e. the region above the QDs. Therefore, we can assume that no phosphorous intermixing inside the QDs has occurred, and only slightly in the capping region. Instead, it is likely that Sb-for-As exchange reactions between QDs and the GaSb cap took place~\cite{Wang_2002_ExchangeAnion}, thus effectively modifying the Sb content of QDs, as will become clearer later on in this study. This mechanism is usually ascribed to the As-for-Sb anion exchange reactions, where Sb exchanges with As~\cite{Wang_2002_ExchangeAnion}. The overall material redistribution promotes a decrease of the compressive strain (from $-3.2\,\%$ without QDs to $-3.0\,\%$ estimated from Raman shift using a model introduced in Ref.~\cite{Montazeri_Nano2010}, see the next section) and probably also to creation of trap states.
Outside the QD area, the concentrations of As and Sb decrease rapidly, while the P level reaches the level of the initial GaP substrate. Since we have largely excluded As-P intermixing between GaP and the QDs, we assume that all phosphorus is bound to GaP, and therefore the concentration of Ga can be divided into Ga concentration in GaP $C_\mathrm{GaP}$ and in the QD area $C_\mathrm{Ga}^\mathrm{QD}$ as
\begin{eqnarray}
C_\mathrm{Ga}=C_\mathrm{GaP}+C_\mathrm{Ga}^\mathrm{QD}=C_\mathrm{P}+C_\mathrm{Ga}^\mathrm{QD} \,,
\end{eqnarray}
where $C_\mathrm{i}$ for $i \in \{\mathrm{Ga}, \mathrm{P}\}$ is the measured concentration of Ga and P. We assume the composition of our QDs as In$_{1-x}$Ga$_{x}$As$_y$Sb$_{1-y}$, thus, we can calculate the effective concentration in the QD area as
\begin{eqnarray}
x=\frac{C_\mathrm{Ga}^\mathrm{QD}}{C_\mathrm{Ga}^\mathrm{QD}+C_\mathrm{In}^\mathrm{QD}} \,,\qquad
y=\frac{C_\mathrm{As}^\mathrm{QD}}{C_\mathrm{As}^\mathrm{QD}+C_\mathrm{Sb}^\mathrm{QD}} \,. \label{eq:Concentration}
\end{eqnarray} 
Given the assumption that In, As, and Sb occur only in the QD area, we can extract the contents of the aforementioned elements using the above equations from EDX data solely from that region. The values of $x$ and $y$ extracted in that way are the following: $x=89-94\,\%$ and $y=82-92\,\%$, see also inset {$\bf{(d)}$}, and we employ these values in our $\mathbf{k\cdot p}$ calculations. 
We would like to point out that the composition in the QDs is averaged across the QD and GaAs IL region, which means the actual amount of Ga and As in the QDs might be overestimated.


\section{Estimate of hydrostatic strain in the GaAs interlayer}
\label{sec:Raman}
The lattice mismatch between GaAs and GaP of $\sim-3.6\,\%$ is released in our structures due to the subsequent growth of QDs and the Sb-rich top layer. In order to estimate the hydrostatic component of strain in the GaAs IL, room temperature Raman measurements have been performed. They have been obtained using NT-MDT spectrometer with a 100$\times$/0.7$\,$NA long working length objective and a 532$\,$nm laser. A 1800$\,$groove/mm grating has been used for dispersion of the scattered light and a thermoelectrically cooled Si CCD camera was used for detection. The spectra have been recorded in $z(xy)z$ backscattering geometry. The measured signals have been fitted by the sum of 3 Lorentzian curves.
Here, we focus on the Raman signal around 290$\,$cm$^{-1}$, where we expect to see the TO phonon of strained GaAs wells (QWs)~\citep{Esther_Nanotech2013}.

\begin{figure}[h]
\centering
\includegraphics[width=\linewidth]{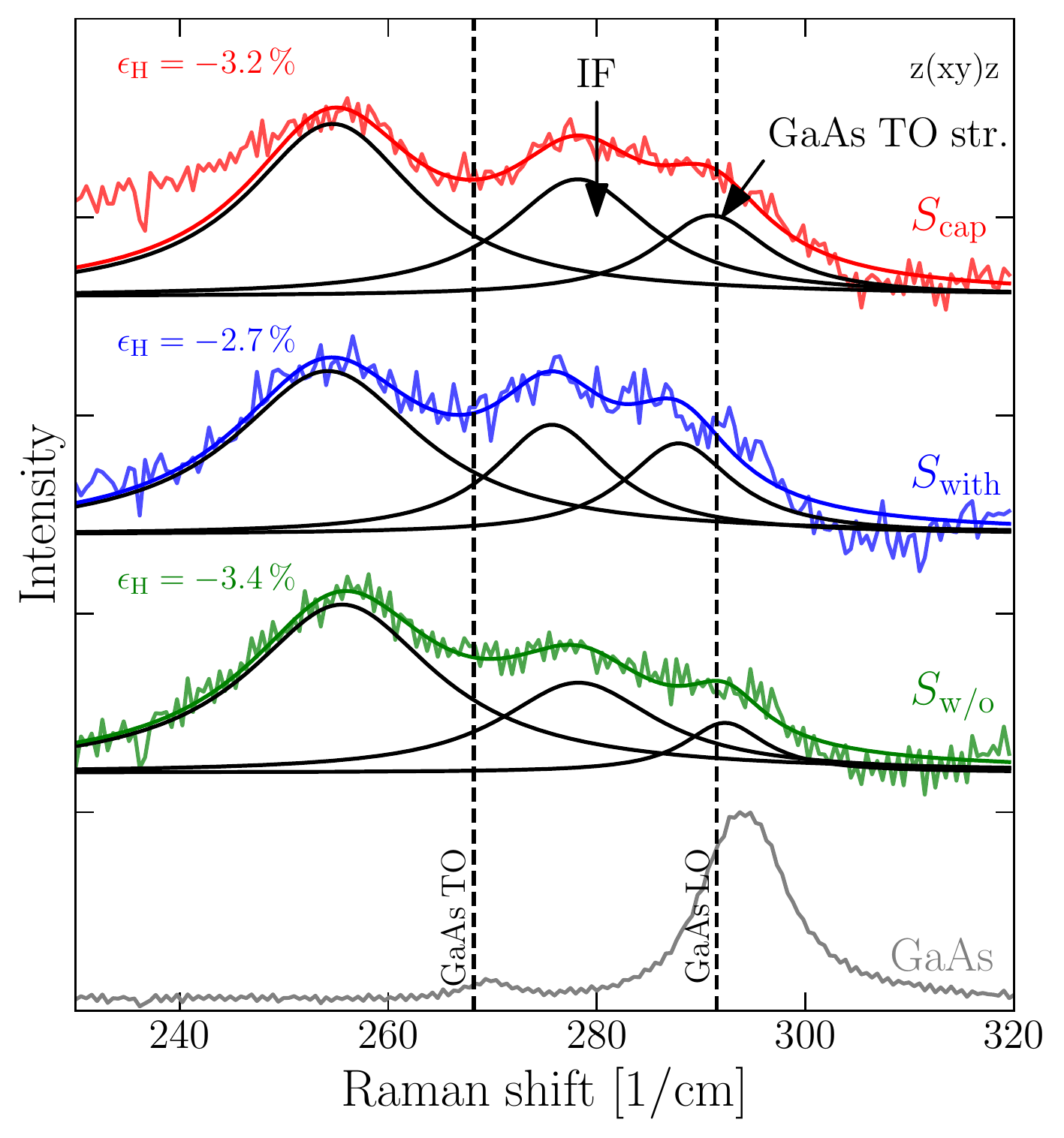}
\caption{Raman spectra taken at room temperature of samples $S_\mathrm{cap}$ (red), $S_\mathrm{with}$ (blue), $S_\mathrm{w/o}$ (green), and bulk GaAs (grey). The dashed lines show the reference bulk GaAs TO and LO phonon modes~\citep{Esther_Nanotech2013}. Calculated hydrostatic strain components $\epsilon_{\mathrm{H}}$ for each sample are also indicated. A label IF corresponds to the interface Raman band.}
\label{fig:raman}
\end{figure}

Fig.~\ref{fig:raman} shows a $\sim19\,\mathrm{cm}^{-1}$ shift of the transverse optical (TO) phonon for the QD and only-GaAs samples, in comparison to GaAs bulk (grey spectrum). By using the strain-dependent \textbf{k$\cdot$p} model studied in Ref.~\cite{Montazeri_Nano2010}, and considering the phonon deformation potentials from Ref.~\cite{Cerdeira_PRB1972}, we can estimate the hydrostatic strain $\epsilon_{\mathrm{H}}$ of the GaAs IL for the three samples studied in this work. We assume the $\sim 19\,\mathrm{cm}^{-1}$ to be the shift of the TO phonon with respect to the bulk value by strain, including approximately $-1\,\mathrm{cm}^{-1}$ correction due to 1D confinement~\cite{NotePCMShift,NotePCM} in the structure, estimated following the Phonon Confinement Model~\cite{Campbell1986,Gouadec2007}. 
By considering $k_\mathrm{TO}$ and $k_\mathrm{TO,B}$ as the Raman shifts of the TO GaAs mode of strained and bulk GaAs, respectively, and $p$ and $q$ the phonon deformation potentials from Ref.~\cite{Cerdeira_PRB1972}, we are able to calculate the hydrostatic strain $\epsilon_{\rm{H}}$ for the GaAs layer in the three different cases. Tab.~\ref{tab:Strain} summarizes the calculated shifts of the TO mode for the investigated samples and for the bulk GaAs/GaP: for $S_\mathrm{w/o}$ it corresponds to $\epsilon_{\mathrm{H}}=-3.4\,\%$, for $S_\mathrm{with}$ to $\epsilon_{\mathrm{H}}=-2.7\,\%$, and for $S_\mathrm{cap}$ we find $\epsilon_{\mathrm{H}}=-3.2\,\%$, which lies in the same order of magnitude of the predicted hydrostatic strain component $\epsilon_{\mathrm{H}}=-3.4\,\%$ of 5\,ML GaAs on GaP by the $\mathbf{k \cdot p}$ calculations. 

%
%
\begin{table}[ht]
\centering
\caption{The estimate of the in-plane strain $\epsilon_{\mathrm{H}}$, compared to the hydrostatic strain between GaAs and GaP bulk (estimated from the bulk lattice mismatch using the relation $\epsilon_{\mathrm{H}}=(a_\mathrm{S}-a)/a$, where the lattice constants of GaAs $a$ and GaP substrate $a_\mathrm{S}$ are taken from Ref.~\cite{Vurgaftman2001}).\vspace{0.5cm}
}\label{tab:Strain}

\begin{tabular}{c|c}
material& $\epsilon_{\mathrm{H}}$ [$\%$]\\
\hline
GaAs/GaP&$ -3.6$ \\
\hline
$S_\mathrm{w/o}$& $-3.4$\\
$S_\mathrm{cap}$& $-3.2$\\
$S_\mathrm{with}$& $-2.7$
\end{tabular}
\end{table}

Through such Raman analysis, we are able to experimentally estimate the variation of the hydrostatic strain $\epsilon_{\rm{H}}$ and to assess the partial relaxation of the strained GaAs layer due to the subsequent growth of QDs. The calculated value $\epsilon_{\rm{H}}$ related to 5\,ML GaAs/GaP agrees very well with the predicted value of $-3.4$ \% calculated for not-disordered GaAs/GaP QW, meaning that we expect a rather abrupt heterostructure interface, with little or no As-P exchange.
The additional growth of the GaSb layer above the QDs is likely to increase of $\epsilon_{\rm{H}}$ in the GaAs IL and thus to reduce the total strain energy, i.e. it acts as a strain compensation layer~\cite{alonsoalvarez_strain_2011}.

\subsection*{Effect of hydrostatic strain on energy levels}

To evaluate the effect of the stain on the energy levels, photoreflectance (PR) was performed at room temperature using the 325\,nm line of HeCd laser (15\,mW) chopped at a frequency of 777\,Hz. The probe beam from a 250 W QTH-lamp was dispersed with a 1/8\,m monochromator. The direct component of the reflected probe beam was recorded with a Si-photodetector.

\begin{figure}[h]
\centering
\includegraphics[width=\linewidth]{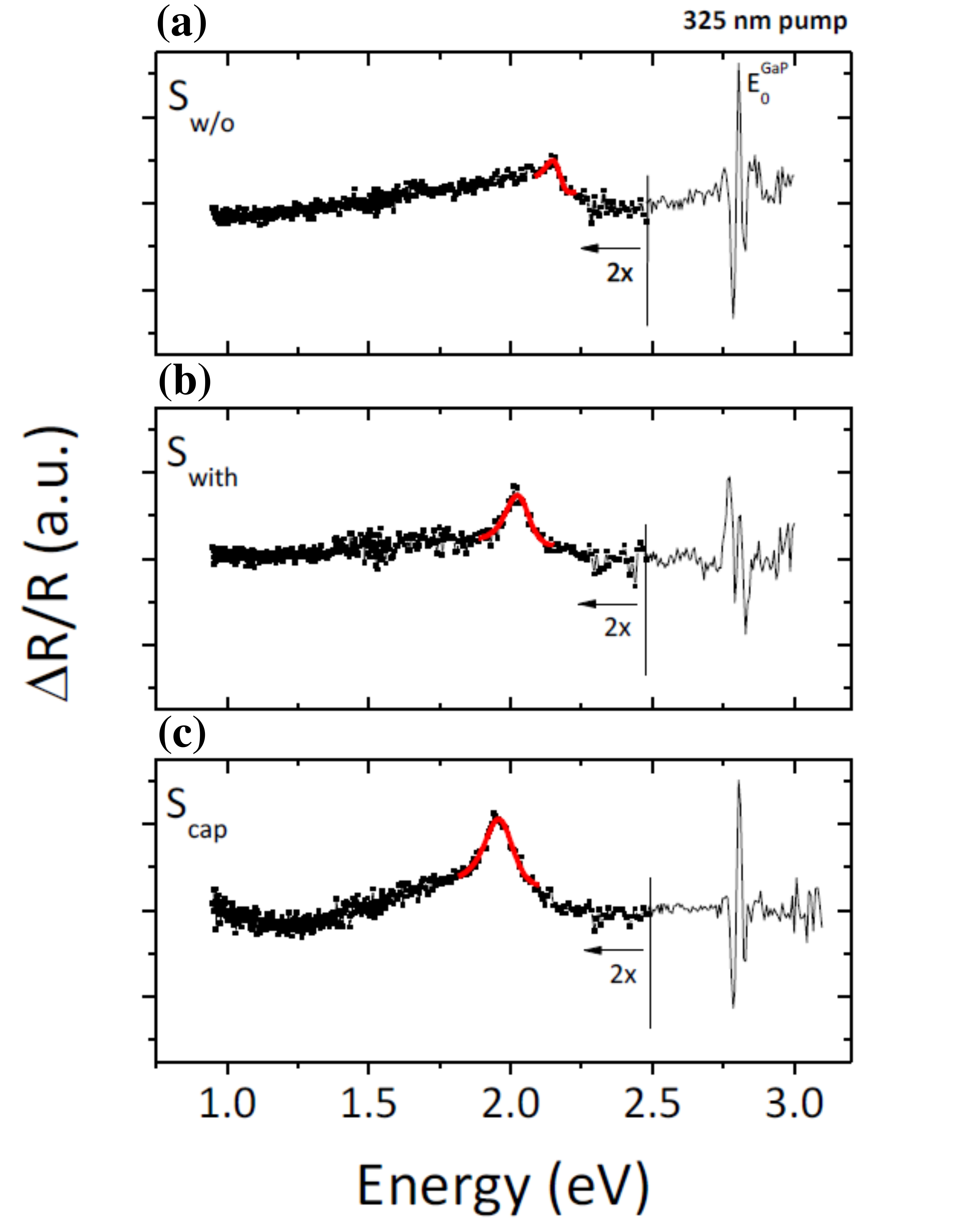}
\caption{PR spectra at room temperature of samples $S_\mathrm{w/o}$, $S_\mathrm{with}$, and $S_\mathrm{cap}$. Solid lines indicate best fits to Aspnes' TDFF~\cite{ASPNES_SurfaceSci73}. Corresponding critical point energies and broadening factors are shown in the Tab.~\ref{tab:PR}.}
\label{fig:PR}
\end{figure}

PR is sensitive only to direct transitions~\cite{Fuertes_2010}. This implies access to transitions involving $\Gamma$-states in the conduction band, however, not the $X$-states of the GaAs/GaP system. The lowest direct (at $\Gamma$ point) critical point of GaP matrix $E_0^\mathrm{GaP}$ is clearly visible in all spectra in agreement with~\cite{Mishima_PR_GaP} around 2.8\,eV (see Fig.~\ref{fig:PR}). Figure~\ref{fig:PR}~(a) shows the results obtained for the sample $S_\mathrm{w/o}$. A single, broad signature is visible around $E_\mathrm{CP}=2.17$\,eV. We assign that signature to the critical point of the GaAs-IL, nominally identical in thickness, but not in strain, in the three samples studied. Indeed the other two samples show similar redshifted broad features as shown in figures~\ref{fig:PR}~(b,c). Given the similarity of the results found for the different samples, we exclude a contribution from the QDs at these energies and relate the observed resonances to the strained IL.
\begin{center}
\begin{table}[ht]
{\small
\hfill{}
\caption{The best fit parameters of Aspnes' TDFF~\cite{ASPNES_SurfaceSci73} with the exponential factor $n=2$ for our samples.}
\vspace{0.5cm}
\begin{ruledtabular}
\begin{tabular}{lcc}
		sample&$E_\mathrm{CP}$ [eV]& broadening [meV]\\ \hline
		$S_\mathrm{w/o}$ & 2.17 & 49\\
		$S_\mathrm{with}$ & 2.03 & 81\\
		$S_\mathrm{cap}$ & 1.97 & 98
\end{tabular} \label{tab:PR}
\end{ruledtabular}}
\hfill{}
\end{table}
\end{center}
Fits to Aspnes' third derivative functional form (TDFF)~\cite{ASPNES_SurfaceSci73} of GaAs-IL signatures yielded critical point energies and broadening factors, as shown in the table~\ref{tab:PR}. The exponential m-factor used (n=2) for best fits is in agreement with confined electronic states within the IL.

The large broadening factors observed ($>50$\,meV) suggest an overlap of transitions involving hh and lh valence band states at RT leading to a single signature and hinder any attempt to the detailed resolution.
In order to compare our PR results from $S_\mathrm{w/o}$ at room temperature with those based on electro-reflectance of Prieto~\textit{et al.} at 80\,K~\cite{Prieto_APL1997}, we assume that the temperature dependence of the GaAs fundamental gap is valid for the narrow IL. We expect thus a rigid shift of 85\,meV due to the 220\,K temperature difference. Applying the same shift to the energy of the observed transition, we find a good agreement between the sample $S_\mathrm{w/o}$ projected to 80\,K and the corresponding sample from Prieto \textit{et al.} (5\,ML thick GaAs/GaP QW), thus confirming the assignment of the PR signature to the GaAs-IL. Two factors are expected to contribute to the differences found between the samples, the energy shift from 1.42\,eV of unstrained bulk GaAs to the range 2.0--2.2\,eV observed for the fundamental direct transition of the GaAs-IL, being based on confinement and strain. Considering the hydrostatic strains $\epsilon_\mathrm{H}$ of the samples, ranging from $-2.7$\,\% up to $-3.4$\,\%, 
and the reported values of conduction- and valence band deformation potential for GaAs as given by Vurgaftman~\textit{et al.}~\cite{Vurgaftman2001}, namely $a_c=-7.17$\,eV and $a_v=-1.16$\,eV, we can estimate the contribution of strain to the energy shift $E_\mathrm{CP}$ of the critical point as
\begin{eqnarray}
\Delta E_\mathrm{CP}=(a_c+a_v)\epsilon_\mathrm{H}\sim 280\,\mathrm{meV}\,.
\end{eqnarray}

This value represents about 48\%
of the observed bandgap shift from unstrained bulk GaAs at 1.42\,eV up to 2.0\,eV, indicating similar contributions of strain and confinement (the remaining 
52\%) to the bandgap widening).
Actually, the maximum expected bandgap variation among samples at the $\Gamma$-point solely due to differences in hydrostatic strain (between $S_\mathrm{w/o}$ and $S_\mathrm{cap}$ samples) can be calculated as
\begin{eqnarray}
\Delta E_\mathrm{CP}=(a_c+a_v)(\epsilon_\mathrm{H}^\mathrm{max}-\epsilon_\mathrm{H}^\mathrm{min}) \sim 50\,\mathrm{meV}\,,
\end{eqnarray}
while the maximum observed shift in $E_\mathrm{CP}$ from PR measurements (see table above) is just 20\,meV (between samples $S_\mathrm{w/o}$ and $S_\mathrm{cap}$), with broadening factors nearly doubling.

This analysis clearly shows that the observed energy shifts of the emission from our samples cannot be solely attributed to differences in accumulated strain in the GaAs-IL. Most likely, the observed shift is the combined effect of strain and confinement variation among samples and is studied in more detail by $\mathrm{{\bf k}}\cdot{ \mathrm{\bf p}}$ simulations in the next section.

\section{$\mathrm{{\bf k}}\cdot{ \mathrm{\bf p}}$ simulations}
To study the origin of the radiative transitions of our samples, calculations based on the combination of one- and eight-band $\mathrm{{\bf k}}\cdot{ \mathrm{\bf p}}$ approximation have been carried out. We consider an In$_{1-x}$Ga$_x$As$_{y}$Sb$_{1-y}$ QD of truncated-pyramid shape, having the dimensions taken from the previous experimental investigations~\cite{Sala2018,t_sala}. 
This QD is then placed on a 5\,ML-thick GaAs IL and embedded in a GaP (001) matrix.

\begin{figure}[h]
	\centering
	\includegraphics[width=1\linewidth]{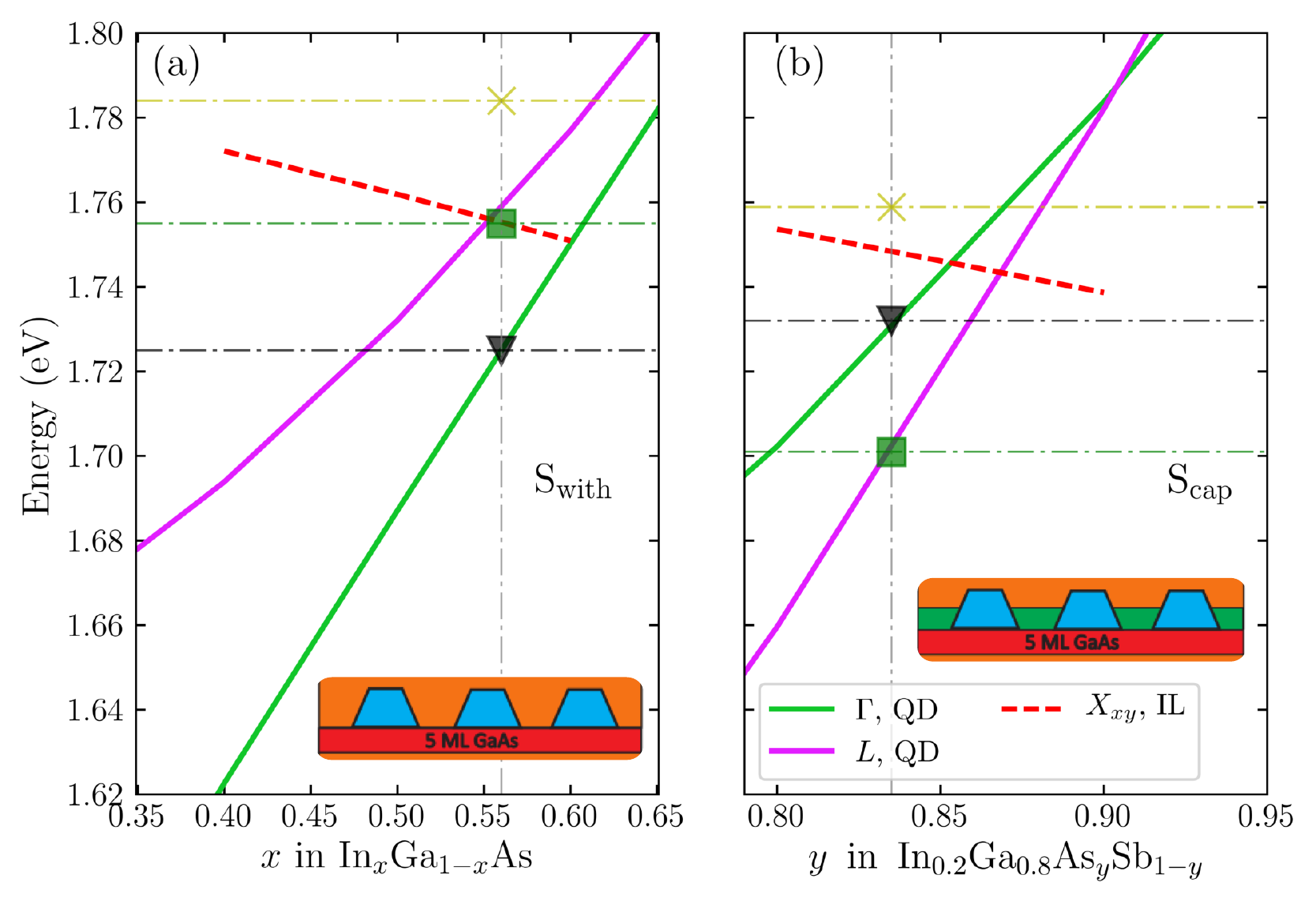}
	\caption{Comparison of the experimentally obtained transition energies for the excitation density $D=0.1\,\mathrm{W/cm^2}$, for samples $S_\mathrm{with}$ (a) and $S_\mathrm{cap}$ (b) (symbols). The corresponding theoretical values for transitions from $\Gamma$ and $L$ electrons (located in QD) to $\Gamma$ holes, obtained by $\mathrm{{\bf k}}\cdot{ \mathrm{\bf p}}$ approximations (solid lines), are also displayed. The red dashed lines represent the transitions energies between $X_{xy}$ electrons and $\Gamma$ holes (both quasiparticles located in IL), extracted from band-edges. The slash-dot vertical lines indicate the concentrations where theory matches the experimental data. Sketches of the corresponding QD areas are also depicted for reader's guidance.}
	\label{fig:kp_results}
\end{figure}

The computational routine started by obtaining the strain field in the whole simulation space using the minimization of the strain energy. The effect of the resulting strain on the confinement potential was then treated using the Bir-Pikus Hamiltonian~\cite{Bir:74} with positionally dependent parameters. The next step involved the self-consistent solution of single-particle Schr\"{o}dinger and Poisson equations, including the effect of piezoelectric fields. All the preceding steps of the calculation were done using the nextnano$++$ simulation suite~\cite{Birner:07,t_zibold}. For more details about our calculation method, we refer to our recent work \cite{Klenovsky2018_TUB}. In the calculations, As and Ga contents in the QDs were varied in intervals close to the experimentally obtained values, e.g.~Ga content in In$_{1-x}$Ga$_x$As QD with respect to the emission of \textit{S$_\mathrm{with}$} and As content in In$_{0.2}$Ga$_{0.8}$As$_{y}$Sb$_{1-y}$ QD to \textit{S$_\mathrm{cap}$}. The magnitude of the valence-band off-set (VBO) of novel heterostructures is usually not well-known. For our calculations we relied on an experimental result, namely the previously measured hole localization energy of InGa(As,Sb) QDs~\cite{Sala2018,t_sala}. This value represents the energy difference between the hole ground state of such QDs and the valence band edge of the surrounding GaP matrix and amounts to 0.370  ($\pm$0.008)\,eV~\cite{Sala2018,t_sala}. Thus, we selected for our calculations a VBO input value of 0.380\,eV. Elsewhere, VBO values for novel heterostructures as for Ga(As,P)/GaP and InAs/AlAs QDs were determined by comparing the calculations of optical transitions with the emission energy in PL investigations~\cite{Shamirzaev_PRB2008,Abramkin_JAP2012}.

The single-particle transition energies of $\Gamma$- and $L$-electrons in the QDs and $X_{xy}$-electrons to $\Gamma$-holes in IL are taken from the band-edges (see panel (b) in Figs.~\ref{pic:Swith_int} and \ref{pic:Scap_int}) and shown in Fig.~\ref{fig:kp_results}. We first notice that states involving $L$-electrons are almost degenerate in energy, hence, we do not distinguish between them in Fig.~\ref{fig:kp_results} and group them under the label $L$.
The same holds true for ($X_{[100]}$, $X_{[010]}$) electrons in the GaAs layer, which we denote $X_{xy}$ (the $X$ bands for GaAs strained to GaP are split into $X_z$ and $X_{xy}$ where $z$ indicates growth direction). 
Note that in QDs, the transitions involving $X$-electrons, even though their eigenvalues are smaller~\cite{Klenovsky2018_TUB} than electrons from $\Gamma$ and $L$, have not been observed in both power- and temperature-dependent PL measurements in~\cite{t_sala}, as well as in our measurements (see also supplementary material, SIII).  
This motivated us to focus only on the spectral range around 1.8\,eV. The weaker oscillator strength of $X$-states in QDs can be understood in the context of Eq.~(\ref{Eq:KaneParamIndirect}) where its weakness is dictated mainly by the indirect origin of electrons (small electron-phonon interaction matrix element). In comparison, $L$ states have, due to intermixing with $\Gamma$-states, much bigger optical coupling making them observable in our PL experiments in the next sections.

\section{Experimental setup for photoluminescence measurements}
The PL measurements have been performed with the samples positioned in a cryostat at 15$\,$K and excited by a continuous wave laser diode having a wavelength of 405$\,$nm and focused to a 0.05$\,$mm$^2$ large spot size. The emitted signal was dispersed by a~1200$\,$grooves/mm ruled grating designed for the wavelength of 750$\,$nm and detected by a photomultiplier connected to a lock-in amplifier. We have performed the following PL experiments: \textit{(i)} excitation density-dependent measurements, where the laser power has been varied by a neutral density (ND) filter across more than 4 orders of magnitude, \textit{(ii)} temperature resolved, where the temperature has been varied from 15$\,$K to room temperature. For the integration time used (0.3$\,$s per wavelength) we have not detected any reasonable PL signal at 300$\,$K which can be continuously fitted by different bands in order to retain physical meaningful information. Therefore, we present here only data up to 100$\,$K which show a reasonable signal for the following analysis. Additionally, the polarization of the PL emission at 15$\,$K has been analyzed by a rotating achromatic half-wave retarder followed by a fixed linear polarizer.

\section{Photoluminescence measurements}
\begin{figure}
	\centering
	\includegraphics[width=1\linewidth]{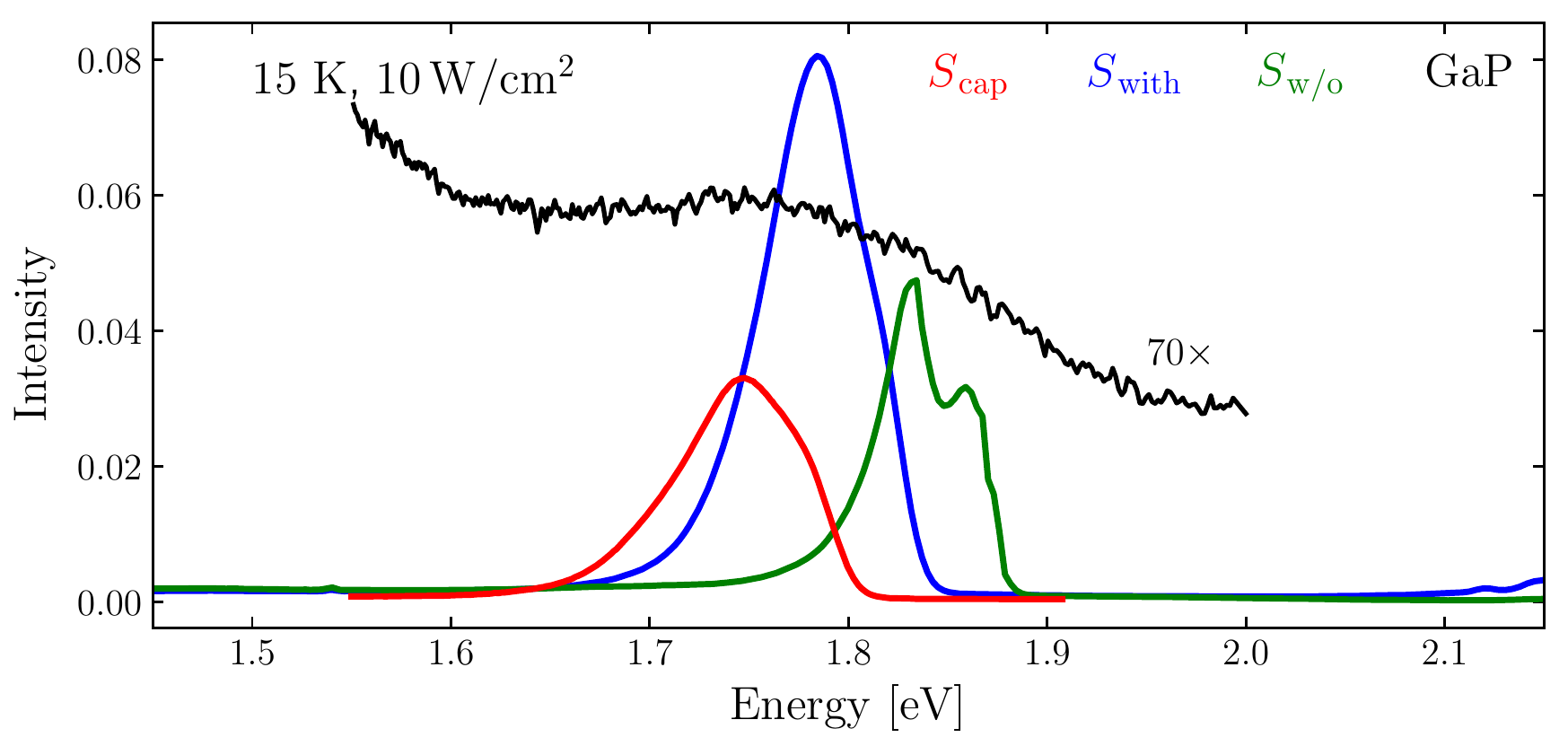} 
	\caption{PL of samples $S_\mathrm{w/o}$ (green curve), $S_\mathrm{with}$ (blue) and $S_\mathrm{cap}$ (red) measured at 15$\,$K, and excitation density of 10$\,$W/cm$^2$. As reference, the background emission from GaP substrate is also shown (black curve) and multiplied by a factor of 70, to facilitate comparison.}
	\label{fig:PL_homogenity}
\end{figure}

Fig.~\ref{fig:PL_homogenity} shows the PL spectra of the three samples centered at the energy of 1.8$\,$eV. The black curve was measured on the bottom of a plain GaP substrate and shows a rather broad background emission, originating from the GaP bulk matrix, which is approximately 70~times weaker than the PL signals detected for the other samples. The two bands in the black spectrum, with energies around 1.8\,eV and between 1.3\,eV and 1.5\,eV, were independently observed during all our measurements and we ascribe them to the emission of donor-acceptor (D,A) pairs (also denoted as DAP) in GaP ~\cite{Dean_PR68,Dean_1970}, see also in the supplementary material (section SIV and reference~\cite{Wight_1968} therein) for more information. Because the intensity of this band is very weak, it is neglected in the analysis of the photoluminescence of our samples.

The PL of sample $S_\mathrm{w/o}$ shows two maxima for 1.83 and 1.86$\,$eV, respectively, while the signal of samples with QDs is shifted to smaller energies: the maxima are located at 1.78$\,$eV for $S_\mathrm{with}$ and 1.74$\,$eV for $S_\mathrm{cap}$, in agreement with previous observations \cite{t_sala}. Finally, we note that the oscillator strength of PL from $S_\mathrm{with}$ is twice larger compared to other samples, which is due to the contribution of the QDs to the PL emission. As we discuss in the supplementary material, all spectra show PL emissions energetically close to each other, around 1.8\,eV. Based on previous experiments on this material system~\cite{Sala2018, t_sala}, it has been shown that such QDs are optically and electrically active, and a significant contribution of the GaAs IL to the PL emission of QDs has to be taken into account. This will be considered in our analysis of our experimental data and in the development of the related theoretical explanation.

\subsection{Sample without QDs $\mathbf{S_\mathrm{w/o}}$}

\begin{figure}
	\centering
	\includegraphics[width=1\linewidth]{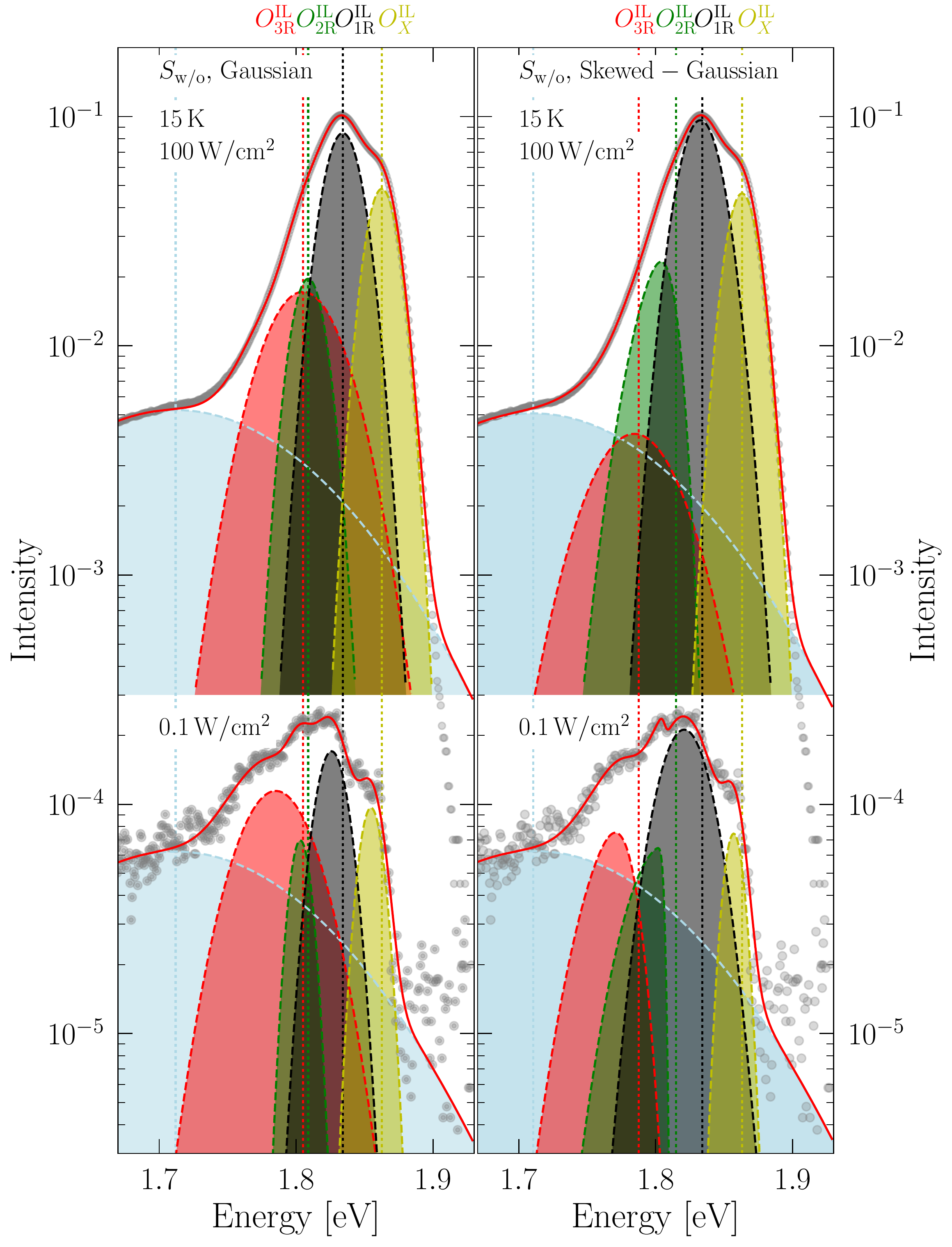}
	\caption{PL spectra of $S_\mathrm{w/o}$ for excitation densities $D$ of 100 and $0.1\,\mathrm{W/cm^2}$ (symbols) and fits (red solid curves) by Eqs.~(\ref{eq:S_wo_G}) (left column) and~(\ref{eq:S_wo_SG}) (right column). Individual transitions are highlighted in different colours. The vertical dashed lines indicate the energy positions of the observed energy transitions at $100\,\mathrm{W/cm^2}$ (see labels of individual bands on the top.}
	\label{fig:S_w/o_intensity}
\end{figure}

Fig.~\ref{fig:S_w/o_intensity} shows the PL emission of sample $S_\mathrm{w/o}$, centered around 1.8 eV. Such a spectrum can be fitted with four emission bands, by using Gaussian curves: from $O_X^\mathrm{IL}$ to $O_\mathrm{3R}^\mathrm{IL}$, labeled from greater to smaller mean emission energy. Such transitions can be ascribed to electrons in the $X_{xy}$ GaAs IL minima recombining with heavy holes in the $\Gamma$ valence band of the GaAs IL~\cite{Prieto_APL1997}. A very similar GaAs/GaP layered system was studied by Prieto \textit{et al.}~\citep{Prieto_APL1997} who reported on optical and theoretical studies on similar structures, such as GaAs/GaP quantum wells (QWs). The authors investigated the optical emission of GaAs QWs embedded in a GaP matrix, with varying thicknesses (between 1 and 6 ML), and then compared the experimental results with calculated ones. They observed that the emission energy of the GaAs layers increased with decreasing layer thickness. Moreover, they noticed that the spectrum consisted of three energy bands separated by 12 and 32$\,$meV, respectively, and that such energy separations mostly do not change with the layer thickness. 
On the contrary, in our fitting routine we were able to fit the GaAs IL emission with one more band, $O_\mathrm{2R}^\mathrm{IL}$, not considered by Prieto \textit{et al.} in their analysis, probably due to their poorer resolution. However, by comparing the fits proposed by Prieto \textit{et al.} (sum of three Gaussians) we have found similar energy separations of the bands, i.e., 12--17$\,$meV and 40--46$\,$meV, depending on the excitation energy used. Moreover, we confirmed their observations, stating that the detected bands cannot be attributed to layer thickness fluctuations, but instead can be referred to phonon-assisted transitions. The corresponding energies closely correspond to TA and LA phonons in GaP~\citep{Prieto_APL1997} and GaAs, as shown in the following table:

\begin{table}[ht]
\centering
\caption{Experimental phonon frequencies (expressed in $\mathrm{cm^{-1}}$) at the high-symmetry points $\Gamma$, $L$ and $X$ of bulk GaP~\cite{GaP_phonons} and GaAs~\cite{Dorner_JAP_GaAsphononsExp}. Parentheses denote ab initio values from Ref.~\cite{Giannozzi_PRB1991_abanitionPhonons}. We present only phonons whose frequencies are closed to values obtained from our PL.}
\vspace{0.5cm}
\begin{tabular}{ccc}
		 & GaP&  GaAs \\ 
\hline 
		$\Gamma_\mathrm{TO}$& 365 &\\
		$L_\mathrm{LA}$& 215 &207 (210)\\
		$L_\mathrm{LO}$& 375 &238 (238)\\
		$L_\mathrm{TO}$& 355 &\\
		$X_\mathrm{LA}$& 249 &225 (223)\\
		$X_\mathrm{TA}$& 104 &82 (82)\\
		$X_\mathrm{LO}$& 370 &\\
		$X_\mathrm{TO}$& 353 &\\
\end{tabular} 
\label{tab:Phon_fre}
\end{table} 

In order to obtain a more precise description of the emission involving also phonon-replicas, an analysis motivated by the line-shape model developed by Christen \textit{et al.} (Eq.~(18a) in Ref.~\cite{CHRISTEN1990}) has been employed. We consider the origin of the broadening to be due to phonons, following the relation for coupling $P_{\mathrm{el}}$ of bulk conduction bands with ${\bf k}\neq 0$ and valence bands at ${\bf k}=0$, which we derived in Ref.~\cite{Klenovsky2018_TUB}

\begin{equation}
\label{Eq:KaneParamIndirect}
P_{\mathrm{el}}\sim(N_{\mathrm{p}}+1)\sum_j\left|\sum_i\frac{\left<u_v^{\Gamma}\left|\mathcal{H}_{\mathrm{eR}}\right|i\right>\left<i\left|\mathcal{H}_{\mathrm{ep}}\right|u_c^{{\bf k}}\right>}{E_{i\Gamma}-E_{\rm ind}-\hbar\omega_j({\bf k})}\right|^2\,,
%
%
%
%
%
\end{equation}
where $i$ and $j$ label the virtual states and the phonon branches for ${\bf k}$, respectively, $u_v^{\Gamma}$ and $u_c^{\bf k}$ mark Bloch waves in ${\bf k}=0$ of valence and ${\bf k}\neq 0$ of conduction band, respectively, $\mathcal{H}_{\mathrm{ep}}$ and $\mathcal{H}_{\mathrm{eR}}$ are hamiltonians for the electron-phonon and electron-photon interaction, respectively, $E_{i\Gamma}$ is the energy of the virtual state in $\Gamma$-point, $E_{\rm ind}$ is the bandgap of the indirect semiconductor, and $\omega_{j}(\bf k)$ marks the frequency of $j$-th phonon branch corresponding to momentum $\bf k$; $\hbar$ marks the reduced Planck's constant. Furthermore, $N_{\mathrm{p}}=\{\exp{[\hbar\omega_{\mathrm{p}}/(k_\mathrm{B}T)]}-1\}^{-1}$ is the Bose-Einstein statistics, where $k_\mathrm{B}$ denotes the Boltzmann constant and $T$ is the temperature. We have inserted the Bose-Einstein statistics into Eq.~(18a) of \cite{CHRISTEN1990} and, thus obtained the relation for Skewed Gaussian profiles of the energy bands, which are assigned being phonon replicas of the zero-phonon line~(ZPL) Gaussian band describing $O_X^\mathrm{IL}$, emitting at 1.857 eV, which we assume to be broadened mainly inhomogeneously due to structural and chemical fluctuations. The same broadening is thus expected also for phonon-replicas which are, however, broadened furthermore due to interaction with crystal lattice via phonons. Hence, the model reads

\begin{widetext}
\begin{eqnarray}
\label{eq:nofix1G3S}
I_{\mathrm SG } = f_\mathrm{G}(I_{\mathrm{G}_i}, E_{\mathrm{G}}, \sigma_{\mathrm{G}_i}) +  \sum_{i=1}^3 f_\mathrm{G}(I_{\mathrm{G}_i}, E_{\mathrm{G}}-E_{{\mathrm {phon}}_i}, \sigma_{\mathrm{G}_i})\cdot \mathrm{erfc}\left(\frac{h\nu -E_{\mathrm G}+E_{\mathrm {phon}_i}}{\sigma_{{\mathrm {phon}}_i}}\right)f_{\mathrm B-E}(E_{{\mathrm{ phon}}_i})\,, \label{eq:S_wo_SG}
\end{eqnarray}
\end{widetext}

where the Gaussian line-shape is represented by $f_{\mathrm{G}} (I,E,\sigma)$, $f_\mathrm{B-E}$ is the Bose-Einstein statistic and $E_\mathrm{phon}$, $\sigma_\mathrm{phon}$ are the phonon energy and phonon broadening, respectively. We compare the model to a more common one based on the sum of the same number of the Gaussian bands
\begin{eqnarray}
I_{\mathrm G } = \sum_{i=0}^3 f_\mathrm{G}(I_{\mathrm{G}_i}, E_{\mathrm{G}_i}, \sigma_{\mathrm{G}_i}) \label{eq:S_wo_G}\,.
\end{eqnarray}
The best obtained fits by both aforementioned models for two different excitation densities $D$ are compared in Fig.~\ref{fig:S_w/o_intensity}.

\begin{figure}
	\centering
	\includegraphics[width=1\linewidth]{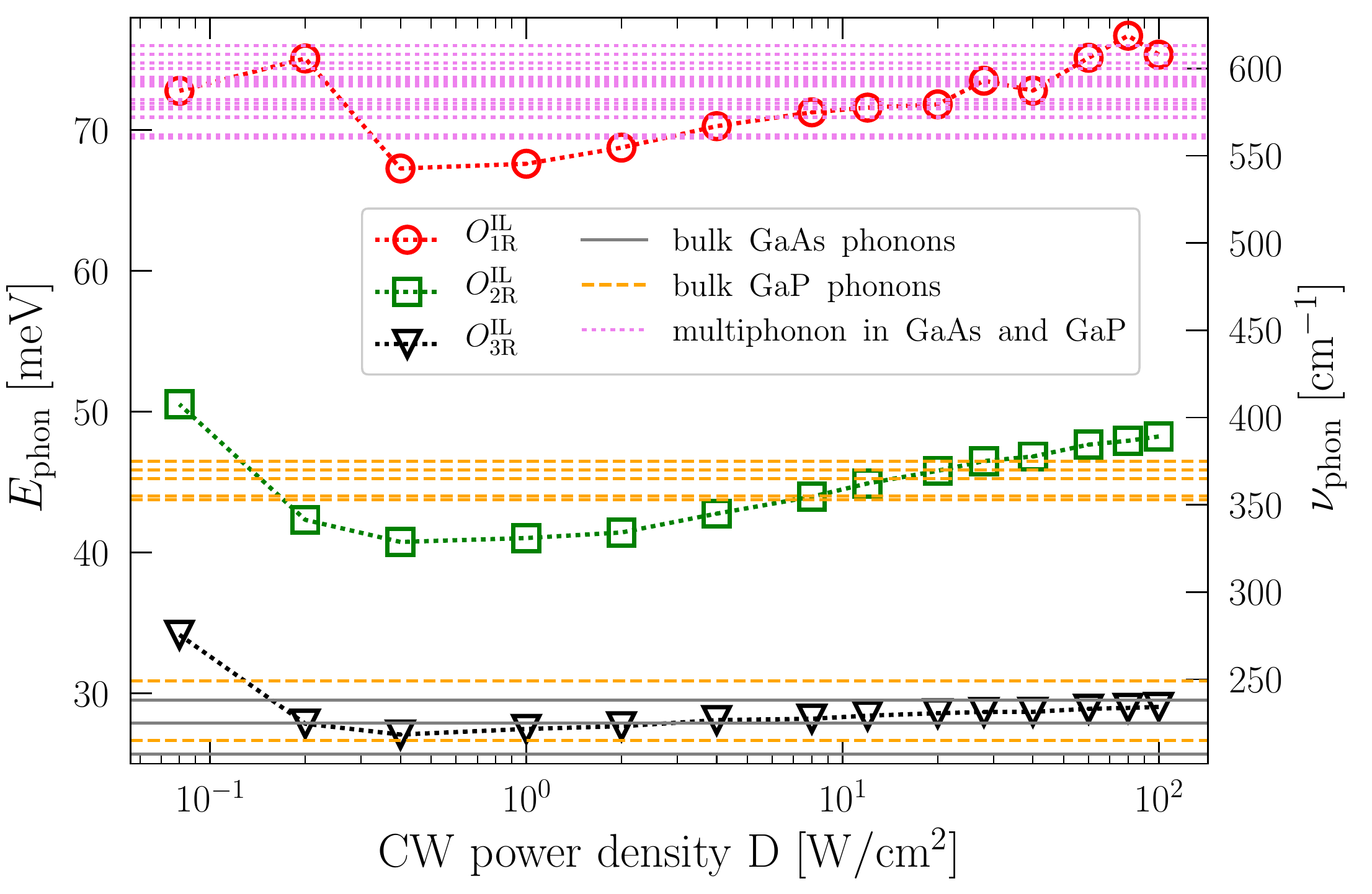}
	\caption{$E_\mathrm{phon}$ for individual phonon-broadened bands from fits by Eq.~(\ref{eq:S_wo_SG}) as a function of $D$ (symbols). Bulk phonon frequencies of GaAs (gray solid line), GaP (dashed orange) listed in Tab.~\ref{tab:Phon_fre}. Violet dotted lines represent multiphonon frequencies from both GaAs and GaP.}
	\label{fig:S_w/o_phonon}
\end{figure}

Based on the model taking into account the phonon-broadening represented by Eq.~(\ref{eq:S_wo_SG}), one ZPL and three phonon-assisted transitions were found (marked by $O_\mathrm{1R}^\mathrm{IL}$ to $O_\mathrm{3R}^\mathrm{IL}$) with phonon energies $E_\mathrm{phon}$ of 30, 45, and 70\,meV (242, 363, 565\,$\mathrm{cm}^{-1}$) (energy differences to the ZPL $O_X^\mathrm{IL}$). The values of $E_\mathrm{phon}$ are spread around frequencies listed in Tab.~\ref{tab:Phon_fre}, see also Fig.~\ref{fig:S_w/o_phonon} for comparison with experiment. The large inhomogeneous broadening of the transitions caused by fluctuation in layer thickness and composition (full width at half maximum $\mathrm{FWHM}$ at minimal excitation density for the bands varies from 20 to 63\,meV, see Tab.~\ref{tab:S_wo_intensity}) does not allow us to determine which particular phonon is involved, and most probably we observe a mix of them, see also Tab.~\ref{tab:Phon_fre}. However, we can at least deduce from Fig.~\ref{fig:S_w/o_phonon} the material to which the phonons belong: for $O_\mathrm{1R}^\mathrm{IL}$ it is GaAs (black), $O_\mathrm{2R}^\mathrm{IL}$ GaP (green) and $O_\mathrm{3R}^\mathrm{IL}$ should belong to a multiphonon recombination both from GaP and GaAs. Note that from the fits of temperature dependence we have found that the energies are higher by $\approx 15$\%, i.e., by $\approx 5$~meV when going from 15~K to 100~K. 

\begin{figure}
	\centering
	\includegraphics[width=1\linewidth]{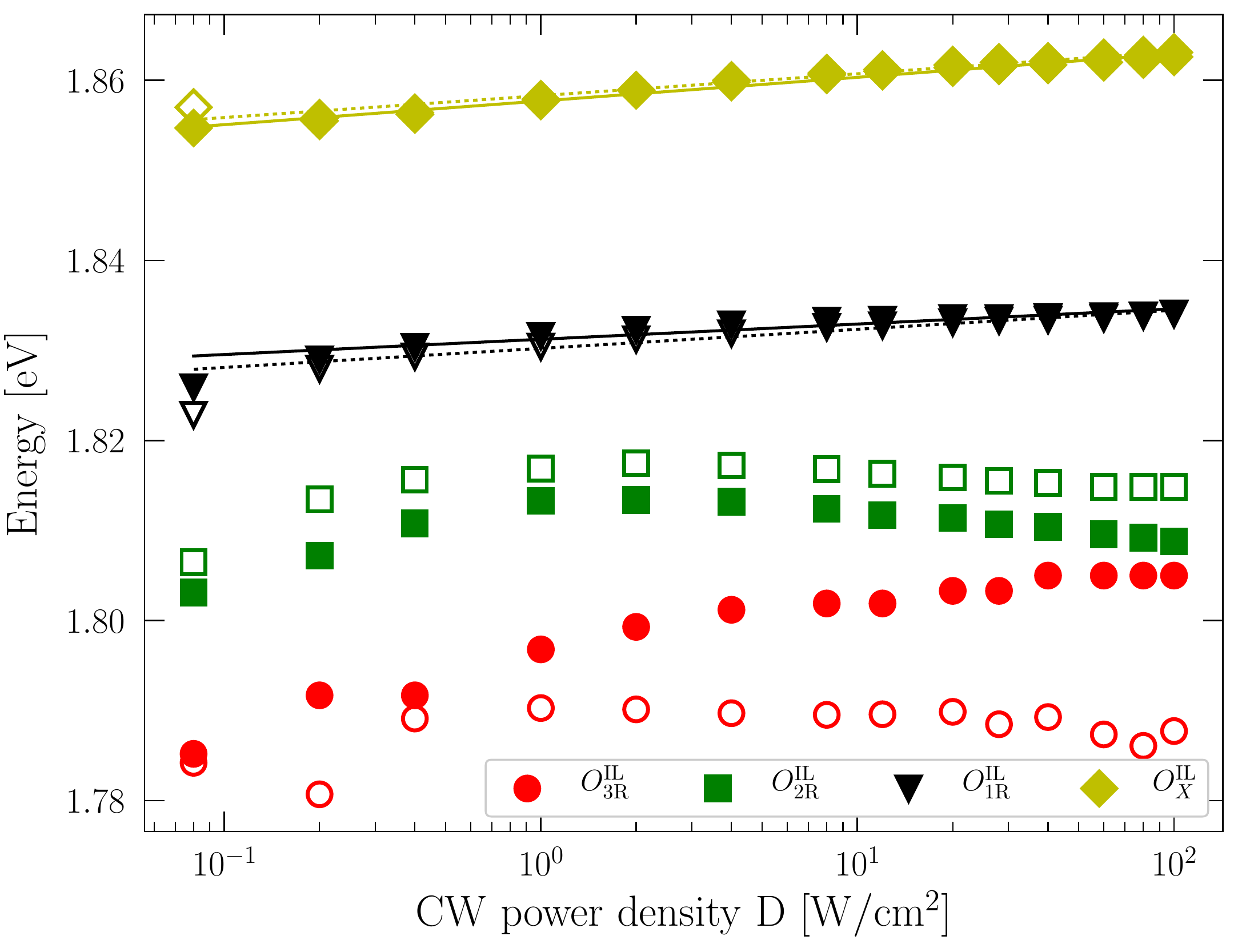}
	\caption{Transition energies from fits by Eq.~(\ref{eq:S_wo_G}) (full symbols) and ~(\ref{eq:S_wo_SG}) (empty symbols) as a function of $D$. Lines represent fits by Abramkim \textit{et al.} model~(\ref{eq:PL_intmodel}). }
	\label{fig:S_w/o_E_int}
\end{figure}

The transition energies and their evolution with excitation density, see Fig.~\ref{fig:S_w/o_E_int}, are very similar for both models. The best-resolved bands $O_X^\mathrm{IL}$ and $O_\mathrm{1R}^\mathrm{IL}$, which are slightly blueshifted with increasing $D$, follow the formula derived by Abramkin \textit{et. al}~\cite{Abramkin_blueshift_analytical} for QWs with a diffused interface, i.e. where the material intermixing leads to fluctuations in QW thickness and alloy composition. That model allows us to distinguish state filling (in both type-I and type-II) and band-bending effects (only in type-II band-alignment structures)~\cite{Abramkin_blueshift_analytical}
\begin{equation}
E(D)=E_\mathrm{I}+\left(U_\mathrm{e}+U_\mathrm{h}\right) \ln\left(D \right)+\beta D^{1/3}, \label{eq:PL_intmodel}
\end{equation}
where $ E_\mathrm {I} $ is extrapolation energy for $ D = 0 \, \mathrm{W / cm ^ 2} $, $ U_\mathrm{e} $ ($ U_\mathrm{h} $) is the electron (hole) Urbach energy tail, and $ \beta $ is the band-bending parameter, respectively. The remaining bands are not monotonous, hence, we do not use Eq.~(\ref{eq:PL_intmodel}) to describe them.

\begin{center}
\begin{table*}[ht]
{\small
\hfill{}
\caption{Summary of the fitting parameters of power density dependent PL for sample $S_\mathrm{w/o}$ obtained by the models ~(G) (Eqs.~(\ref{eq:S_wo_G})) and ~(SG) (\ref{eq:S_wo_SG}) and fit by Eq.~(\ref{eq:PL_intmodel}) with $E_\mathrm{I}=E(D=0)$ and Urbach energy tail $(U_\mathrm{e}+U_\mathrm{h})$. Values of FWHM and emission energies $E$ marked by $^*$ were obtained for $D=0.1\,\mathrm{W/cm^2}$. Exponents $\gamma^{\pm \mathrm{error}}$ are sorted by region as follows: $\gamma_A/\gamma_B/\gamma_C$, see text. Phonon-assisted transitions are labeled using ``rep.".}
\vspace{0.5cm}
\begin{ruledtabular}
\begin{tabular}{lccccccc}
		transition & model & $^*$FWHM [meV]&  $^{*}E$ [meV]& $E_\mathrm{I}$ [meV]&  $U_\mathrm{e}+U_\mathrm{h}$ [meV]   
		& $\gamma$& band alignment \\ 
\hline 
        $O_X^\mathrm{IL}$ & G& 21 &  1855 &$1847\pm1$ & $1.1\pm0.1$ &  $1.53^{\pm0.07}/0.66^{\pm0.02}/0.24^{\pm0.03}$& Type-I \\ 
		$O_\mathrm{1R}^\mathrm{IL}$& G &27 &  1826 &$1824\pm1$ & $0.7\pm0.1$ &  $1.28^{\pm0.04}/0.76^{\pm0.03}/0.41^{\pm0.03}$& Type-I \\ 
		$O_X^\mathrm{IL}$ & SG & 18 &  1857&$1849\pm1$ & $1.1\pm0.1$ &  $1.7^{\pm0.1}/0.65^{\pm0.02}/0.22^{\pm0.03}$& Type-I \\ 
		$O_\mathrm{1R}^\mathrm{IL}$ & SG &43 &  1823& $1828\pm1$ & $0.9\pm0.1$ &  $1.1^{\pm0.1}/0.74^{\pm0.02}/0.41^{\pm0.03}$ & Type-I\\ 
		$O_\mathrm{2R}^\mathrm{IL}$ & SG & 57 &  1806& &  &  $1.2^{\pm0.3}/0.62^{\pm0.02}/0.42^{\pm0.03}$& Type-I \\ 
		$O_\mathrm{3R}^\mathrm{IL}$ & SG & 63 &   1784& &  &  $1.2^{\pm0.2}/0.50^{\pm 0.01}/0.07^{\pm0.07}$& Type-I
\end{tabular} \label{tab:S_wo_intensity}
\end{ruledtabular}}
\hfill{}
\end{table*}
\end{center}

Based on the fitted values of parameters to the model~(\ref{eq:PL_intmodel}), we have determined the band alignment of $O_X^\mathrm{IL}$ and $O_\mathrm{1R}^\mathrm{IL}$ to be of type-I origin, which is consistent with the energy blueshift described only by the Urbach energy tails ($U_\mathrm{e}+U_\mathrm{h}\approx 1\,$meV). That clearly indicates that the origin of those transitions is spatially-direct electron-hole recombination in IL.

\begin{figure}
	\centering
	\includegraphics[width=1\linewidth]{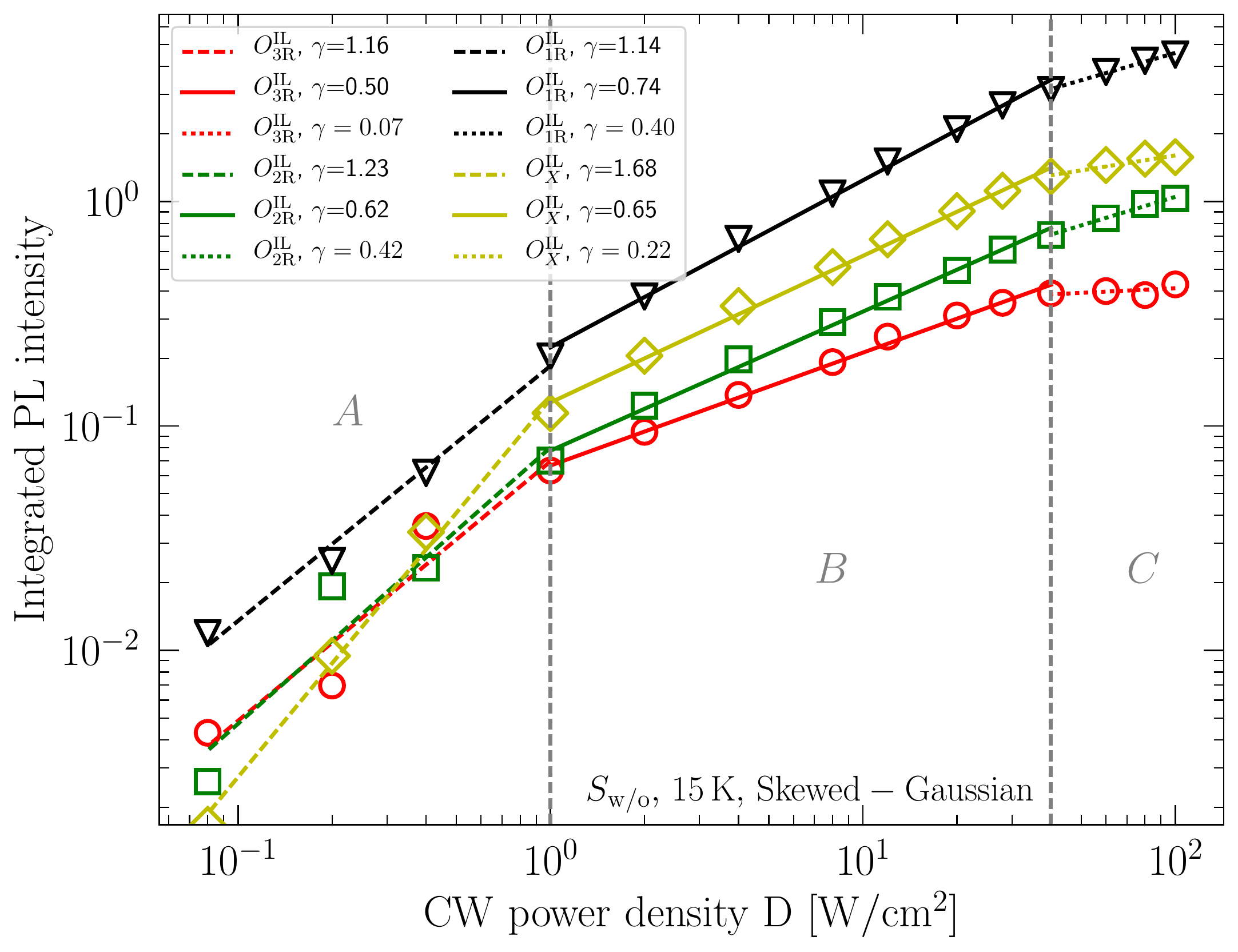}
	\caption{Integrated PL intensity of individual bands obtained by fits by Eq.~(\ref{eq:S_wo_SG}) as a function of excitation density $D$ (symbols) fitted in three regions by $I\propto D^\gamma$. The exponents $\gamma$ are added in legend.}
	\label{fig:S_w/o_Osc}
\end{figure}

In general, the PL intensity $I$ shows a power law dependence on the excitation density $D$,~i.e., $I \propto D^\gamma$, where $\gamma$ represents the mechanism causing the transition. In case of thermodynamic equilibrium between recombination and generation rate, and in the low excitation power regime (when the Auger processes can be neglected), the equations (25)--(27) of Ref.~\cite{Schmidt_PRB1992} showing rate proportionality to $D$ developed by Schmidt~{\textit{et al.}}~\cite{Schmidt_PRB1992} can be used. Based on that relation, $\gamma$ for individual types of transitions has been determined using the following rules: $\gamma \sim 1$ implies an exciton like transition, $\gamma\sim 2$ biexciton, and $\gamma < 1$ suggests a recombination path involving defects or impurities such as free-to-bound (recombination of a free hole and a neutral donor or of a free electron and a neutral acceptor) and donor-acceptor pair transitions. The integrated intensities of all fitted PL bands of the emission from sample $S_\mathrm{w/o}$ are shown in Fig.~\ref{fig:S_w/o_Osc}. They can be divided into three segments where the dependencies follow a linear behaviour on a log-log scale. In region $A$, an exciton recombination ($1.14<\gamma<1.68$ for all bands) prevails for $D<1\,\mathrm{W/cm^2}$, followed by region $B$ where recombination of free carriers with traps is clearly affecting the emission intensity ($0.5<\gamma<0.74$). For $D>40\,\mathrm{W/cm^2}$, i.e. in the region $C$, we observe the beginning of saturation (traps states becoming fully occupied) with $\gamma<0.4$~\cite{Schmidt_PRB1992}. All fitted parameters of power dependencies of $O_X^\mathrm{IL}$ and $O_\mathrm{1R}^\mathrm{IL}$ of both models~(\ref{eq:S_wo_G}) and~(\ref{eq:S_wo_SG}) are summed up in Tab.~\ref{tab:S_wo_intensity}.

The effect of temperature on band gap energy shrinkage has been quantified through several empirical or semi-empirical models. Among the empirical ones for III-V semiconductors, the Varshni relation~\cite{Varshni} is often used to assess nonlinear temperature dependent bandgap shift, i.e.
\begin{equation}
E(T)=E_{\mathrm{V},0}-\frac{\alpha T^2}{T+\beta_\mathrm{V}}\,, \label{eq:Varshni}
\end{equation}
where $E_{\mathrm{V},0}$ is the emission energy at temperature $0\,\mathrm{K}$, $\alpha$ the Varshni parameters characterizing the considered material, and $\beta_\mathrm{V}$ describes the rate of change of the bandgap with temperature and the frequency, which is a modified Debye one, respectively. The validity and physical significance of the Varshni parameters can be best judged by comparing to another model, e.g., the power-function model of P{\"a}ssler \textit{et~al.}~\cite{Passler_PRB2002, Passler_PSS2002}, which shows that the $\beta_\mathrm{V}$ parameter is connected with the Debye frequency $\Theta_\mathrm{D}$ as $\Theta_\mathrm{D}=2\beta_\mathrm{V}$. 
\begin{center}
\begin{table*}[!ht]
{\small
\hfill{}
\caption{Parameters of the temperature evolution of emission energies in sense of the Varshni parameters $E_{V,0}$, $\alpha$ and $\beta_\mathrm{V}$, and intensities following the Boltzmann model~Eq.~(\ref{eq:Arrhenius}) with two activation energies $E_1$, $E_2$ and corresponding ratio of radiative and nonradiative lifetimes $\tau_0/\tau_1^\mathrm{NR}$, $\tau_0/\tau_2^\mathrm{NR}$, of individual bands of sample $S_\mathrm{w/o}$. For comparison, bulk values of expected materials taken from Ref.~\cite{Vurgaftman2001} are added as well. The accuracy of fitted parameters is better than 3\,\% except values marked by $^*$ with have accuracy $\approx$12\,\%. Phonon-assisted transitions are labeled using ''rep.".}
\begin{ruledtabular}
\begin{tabular}{lccc|cccc}
				transition & $E_{V,0}$ [meV]& $\alpha$ [$10^{-4}\,\mathrm{eVK^{-1}}$]& $\beta_\mathrm{V}$ [K]&   $\tau_0/\tau_1^\mathrm{NR}$ & $E_1$ [meV]& $\tau_0/\tau_2^\mathrm{NR}$ $\times 10^3$ & $E_2$ [meV]\\ 	
		\hline
		$O_X^\mathrm{IL}$ & $1861$ & $1.0^*$ &33.7& $34.7$& $11.5 $& $2.9$& $33.7$\\ 
		$O_\mathrm{1R}^\mathrm{IL}$& $1832$ & $3.21$& $298$ &$43.1$& $12.4 $& $2.1$& $33.2$\\
		$O_\mathrm{2R}^\mathrm{IL}$ & $1809$& $5.10$& $247$& $10.7$& $10.7 $& $2.63$& $33.2$\\ 
		$O_\mathrm{3R}^\mathrm{IL}$ & $1780$& $8.89$& $299$& $27.9 $& $12.1 $& $126$& $50.4$\\ 
		\hline
		GaAs, $\Gamma$ ($L$) [$X$]& 1519 (1815) [1981]& 5.405 (6.05) [4.60]& 204& \\
		GaSb, $\Gamma$ ($L$) [$X$]& 812 (875) [1141]& 4.17 (5.97) [4.75]& 140 [94]& \\
		InAs, $\Gamma$ ($L$)& 417 (1133)& 2.76& 93& \\
		InSb, $\Gamma$ ($L$) [$X$]& 235 (930) [630]& 3.20& 170&\\
		GaP, $\Gamma$ ($L$) [$X$]& 2886 (2720) [2350]& 5.77& 372&
\end{tabular} \label{tab:S_wo_Varshni}
\end{ruledtabular}}
\hfill{}
\end{table*}
\end{center}
Another well-established model evaluates the decrease in the energy thresholds, which are proportional to factors of the Bose-Einstein statistics for phonon emission and absorption~\cite{ODonnell_APL1991,Vina_PRB1984,Lautenschlager_PRB1987}
\begin{equation}
E(T)=E_{\mathrm{B},0}-S  \overline{E}_\mathrm{phon} \left[\coth(\overline{E}_\mathrm{phon}/2k_\mathrm{B}T) -1    \right]\,, \label{eq:Boltzmann}
\end{equation}
where $E_{\mathrm{B},0}$ is the band gap at zero temperature, $S$ is a dimensionless exciton-phonon coupling constant, and $\overline{E}_\mathrm{phon}=\Theta_\mathrm{E}k_\mathrm{B}$ is an average phonon energy related to the Einstein frequency $\Theta_\mathrm{E}$, while $k_\mathrm{B}$ is the Boltzmann constant. Both models can be compared using the values of the Einstein and the Debye frequencies as $\Theta_\mathrm{D}/\Theta_\mathrm{E}=4/3$. Such models were developed for bulk materials but are commonly used to describe the temperature evolution of transition energies of low-dimensional systems and we adopted them for our analysis. The Varshni model overestimates parameters and experimental data at cryogenic temperatures (in our case lower than 30\,K). Because the models are not taking into account any thermalization effects, which can be significant at cryogenic temperatures for many low dimensional structures such as quantum wells~\cite{Grassi_PRB2000}, quantum rings~\cite{Sibirmovsky_2015} and especially QDs~\cite{Dai_JAP1997, Alouane_2014}, we use here for the evaluation of thermalization effects the correction of the Varshni model proposed by Eliseev~\citep{Eliseev_apl2003_PLtemp}.

That model assumes also an impurity effect on excitonic band (created by an intermixing between QD material and surrounding layer) modelled by the Gaussian-type distribution of energy of the localised states with broadening parameter (carrier disorder energy) $\sigma_E$ resulting in a Stokes redshift $-\sigma_E^2/k_\mathrm{B} T$ \cite{Chang_apl2008}
\begin{equation}
E(T)=E_{\mathrm{V},0}-\frac{\alpha T^2}{T+\beta_\mathrm{V}}-\frac{\sigma_E^2}{k_\mathrm{B}T}\,. \label{eq:Varshni-like}
\end{equation}

\begin{figure}[!h]
	\centering
	\includegraphics[width=1\linewidth]{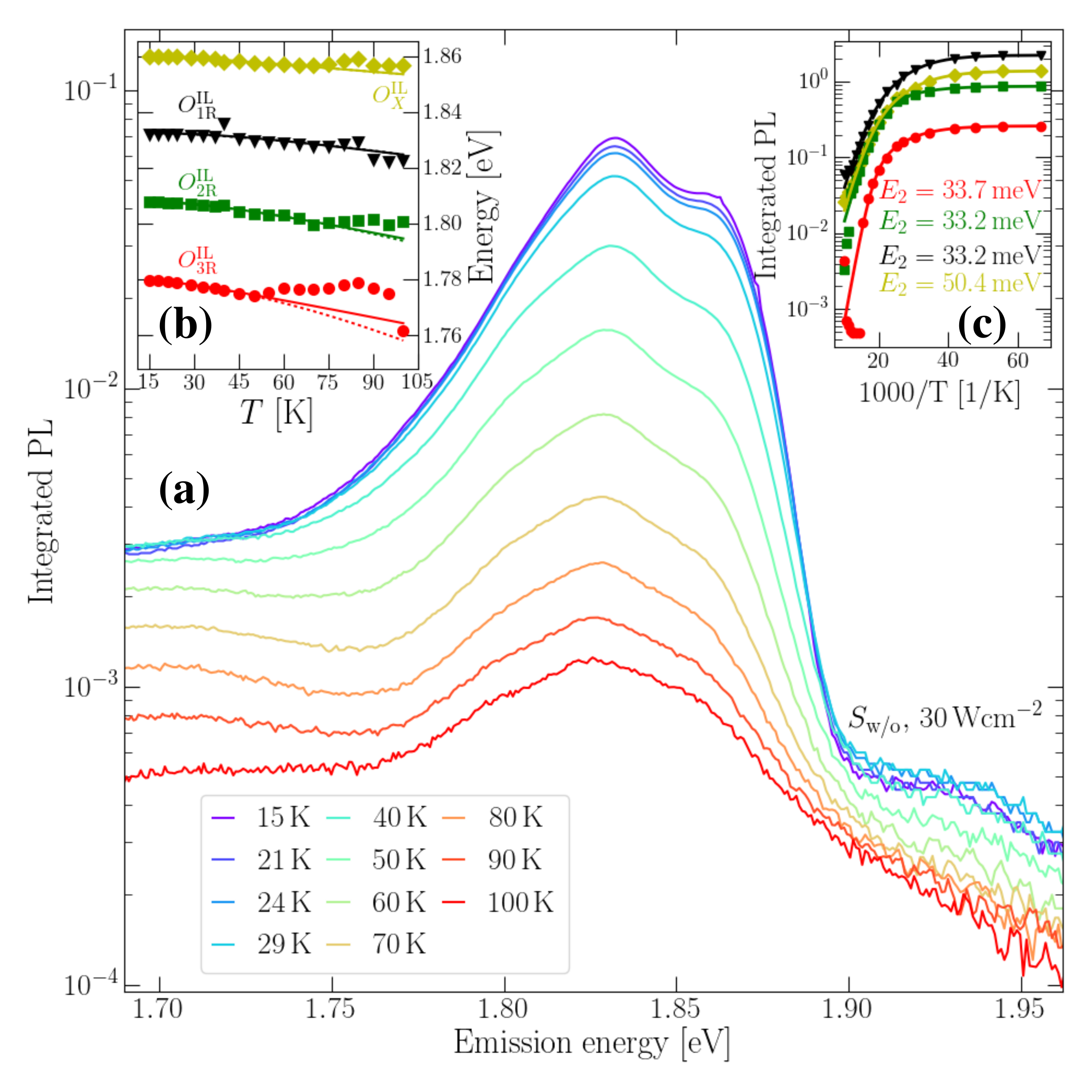}
	\caption{$S_\mathrm{w/o}$: (a) measured PL at $D=30\,\mathrm{Wcm^{-2}}$ between 15\,K and 100\,K. Inset (b) shows the best fit of emission energy of $O_X^\mathrm{IL}$--$O_\mathrm{3R}^\mathrm{IL}$ (symbols) as a function of temperature by the Varshni Eq.~(\ref{eq:Varshni}), solid curve, [modified Varshni Eq.~(\ref{eq:Varshni-like}), dashed curve] model. In the inset (c) we give integrated PL intensities of individual transitions $O_X^\mathrm{IL}$--$O_\mathrm{3R}^\mathrm{IL}$ (symbols) fitted by the Boltzmann model, Eq.~(\ref{eq:Arrhenius}) (solid curves), with the high temperature activation energies $E_2$ shown in the inset.}\label{pic:Swo_temp}
\end{figure}

We employed both aforementioned models to analyze emission energies obtained from fits of temperature resolved PL shown in Fig.~\ref{pic:Swo_temp}~(a) with the corresponding fits by Eq.~(\ref{eq:S_wo_SG}) given in the inset (b). The parameters of the best fits are summarized in Tab.~\ref{tab:S_wo_Varshni}.
Interestingly, the transition from band $X_{xy}$ to heavy holes, $O_X^\mathrm{IL}$, has significantly smaller Varshni parameters compared to unstrained bulk GaAs~\cite{Vurgaftman2001}. On the other hand, the parameters for phonon-assisted transitions $O_\mathrm{1R}^\mathrm{IL}$ and $O_\mathrm{2R}^\mathrm{IL}$ are reasonably close to bulk transition from the $X$ point to the valence band. The coupling-parameter $S$ increases with the increasing total number of phonons available from 0.52 to 1.3. We find average phonon energies between 4.6 and 10.4~meV which can be due to temperature quenching of the PL associated with carrier recombination trough impurities. Most probably, the quenching mechanism is related to nitrogen complexes with activation energy $E_\mathrm{A}=8\,$meV~\cite{ioffe} which is often present during GaP growth~\cite{Skazochkin_GaPtraps}. All parameters related to temperature changes of gap energies can be found in the supplementary material.

%

The mechanisms responsible for the temperature quenching of PL intensity, $I_\mathrm{PL}(T)$, can be accounted for by the Boltzmann model for excitonic recombination with two characteristic activation energies~\citep{Daly_prb1995, Alen_apl2011}
\begin{equation}
I_\mathrm{PL}(T)=\frac{I_0}{1+\tau_0\left[\Gamma_1\exp(-E_1/k_\mathrm{B}T)+\Gamma_2\exp(-E_2/k_\mathrm{B}T)\right]},               \label{eq:Arrhenius}
\end{equation}
where $I_0$ is the intensity at 15$\,$K (lowest temperature reached in our measurements), $\tau_0$ is temperature-independent radiative recombination time at 15$\,$K, $E_1$ and $E_2$ are the activation energies of the two quenching mechanisms with related scattering rates $\Gamma_1$ ($\Gamma_1=1/\tau_1^\mathrm{NR}$) and $\Gamma_2$ ($\Gamma_2=1/\tau_2^\mathrm{NR}$).

That model is employed for various temperatures for sample $S_\mathrm{w/o}$ in the inset Fig.~\ref{pic:Swo_temp}~(c) and we found two similar mechanisms for all bands described by activation energies of impurities (nitrogen and oxygen complexes~\cite{ioffe, Dean_1966_nitrogenInGaP, Dean_1968_oxygenInGaP} or redistribution of material): $E_1$ around 10--12\,meV (80--100\,$\mathrm{cm^{-1}}$) and phonons $E_2= 33\,\mathrm{meV}$ (266\,$\mathrm{cm^{-1}}$) which are comparable with phonon energies in bulk GaAs or GaP listed in Tab.~\ref{tab:Phon_fre}.

\subsection{Sample with QDs $\mathbf{S_\mathrm{with}}$}
\begin{figure}[!h]
	\centering
	\includegraphics[width=1\linewidth]{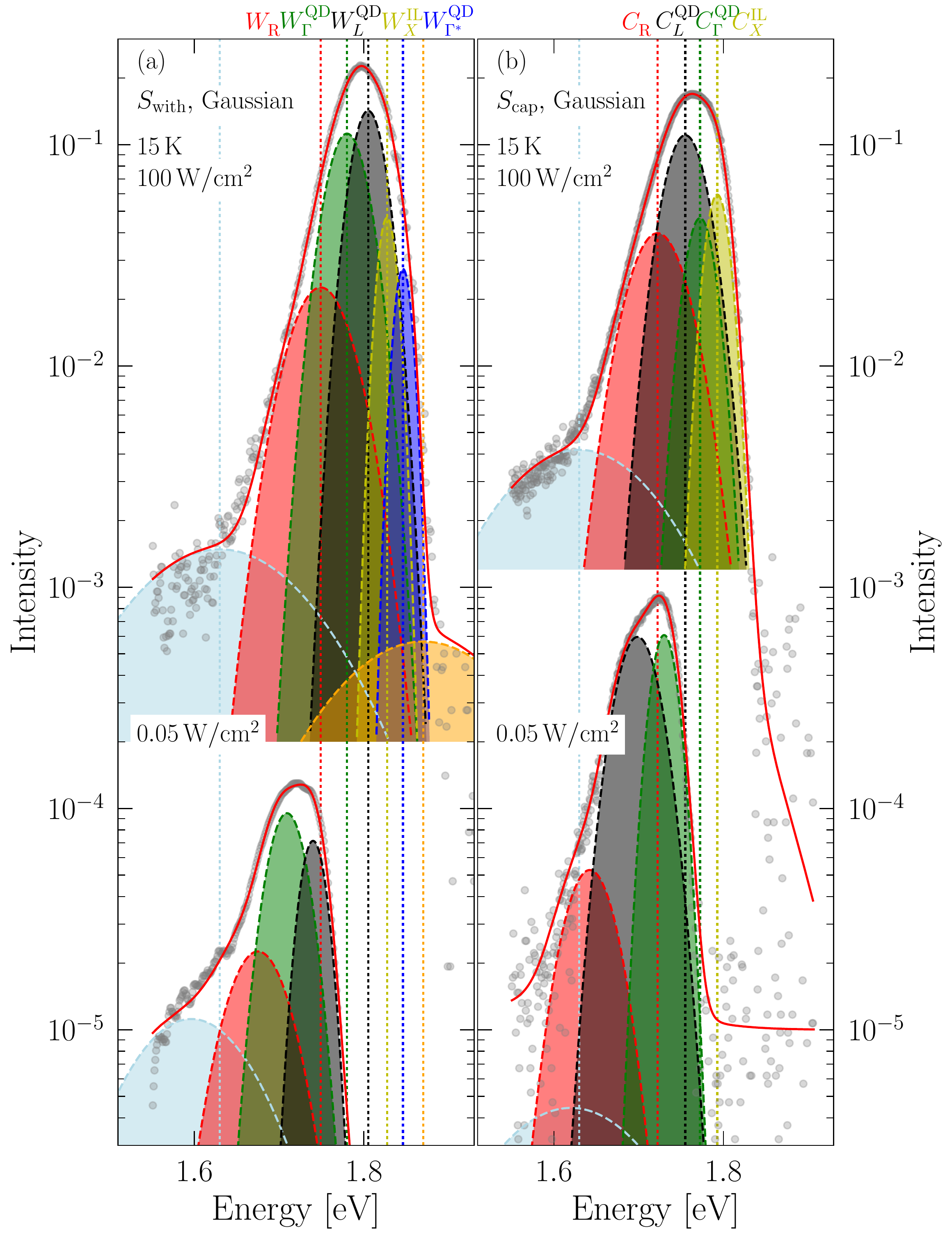}
	\caption{PL spectra (points) of samples $S_\mathrm{with}$ (a) and $S_\mathrm{cap}$ (b) for $D=$100\,W/cm$^2$ and 0.05\,W/cm$^2$ fitted by sum of five Gaussian bands. Individual transitions are shadowed and for estimation of their energy-shift the vertical lines are added in positions of transition energies observed at 100\,W/cm$^2$. Labels of bands are given at the top.}
	\label{fig:PL_int_QDs}
\end{figure}
\label{Sec:PL_int_wo}
PL spectra of the sample ${S_\mathrm{with}}$ as a function of both excitation and temperature dependence were fitted by the 
sum of 5~Gaussian bands, labeled from smaller to greater mean energy as $W_\mathrm{R}$ to $W_{\Gamma^*}^\mathrm{QD}$, see Fig.~\ref{fig:PL_int_QDs}~(a) where fits for two different $D$ values are shown. The physical meaning of the parameters such as the emission energy, FWHM, and PL intensity were investigated using the appropriate aforementioned models and the parameters found are summarized in Tabs.~\ref{tab:QDs_intensity}--\ref{tab:QDs_temperature}. All parameters of the fits are given in the supplement.

\begin{figure*}[!]
	\centering
	\includegraphics[width=1.0\linewidth]{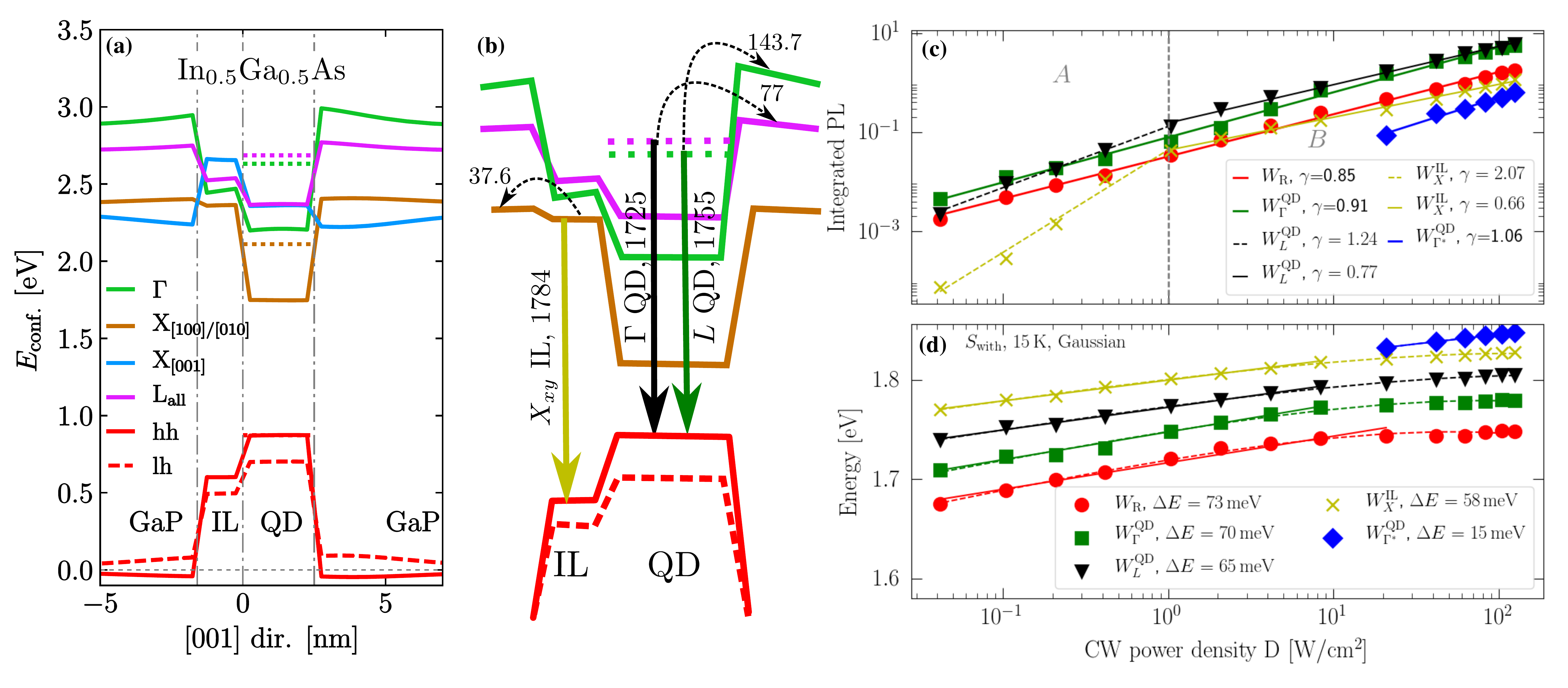}
	\caption{$S_\mathrm{with}$: (a) Bandscheme of $\mathrm{In_{0.5}Ga_{0.5}As}$ QD calculated using nextnano++~\cite{Birner:07,t_zibold} along with single-particle eigenenergies (dotted lines). (b) Detail of the bandscheme with showing also recombination paths (solid arrows with transition energies in meV) and escape energies (values in meV) observed in temperature dependent PL. (c) Integrated PL intensity of individual transitions (symbols) fitted by linear dispersion in log-log scale, corresponding slopes are given in the legend. (d) Emission energy of individual transitions (symbols) vs. $D$; values of absolute blueshifts are in the inset. Solid (dashed) curves represent fits by Eq.~(\ref{eq:PL_intmodel})
	with $\beta=0$ ($\beta\neq0$).}\label{pic:Swith_int}
\end{figure*}

As was menti oned before, the contributions from IL and QDs transitions overlap around 1.8\,eV, which renders the assignment of the Gaussian-like bands to radiative channels more complex. Due to this phenomenon, we observe a much stronger emission from the bands originating from QDs (labeled as $W_{\Gamma}^\mathrm{QD}$ and $W_{L}^\mathrm{QD}$), compared to what one which would be expected by  Eq.~(\ref{Eq:KaneParamIndirect}). A distortion of the effective $\gamma$-parameters of both transitions is concluded.
\begin{center}
\begin{table*}[ht]
{\small
\hfill{}
\caption{Summary of the fitting parameters of power density dependencies of PL for samples $S_\mathrm{with}$ and $S_\mathrm{cap}$. Values of FWHM and emission energies $E$ marked by $^*$ were obtained at $D=0.1\,\mathrm{W/cm^2}$. Energy shift described by Eq.~(\ref{eq:PL_intmodel}) with $E_\mathrm{I}=E(D=0)$, Urbach energy tail $(U_\mathrm{e}+U_\mathrm{h})$ and band-bending parameter $\beta$. Values given in brackets are best-fit parameters using Eq.~(\ref{eq:PL_intmodel}) with $\beta\neq0$. If exponent $\gamma^{\pm \mathrm{error}}$ differs during measured excitation density range, we sort them according to the corresponding region as follows: $\gamma_A/\gamma_B$.}\label{tab:QDs_intensity}
\begin{ruledtabular}
\begin{tabular}{lcccccc}
		transition  & $^*$FWHM [meV]& $^*E$ [meV]& $E_\mathrm{I}$ [meV]&  $U_\mathrm{e}+U_\mathrm{h}$ [meV]  & $\beta$ [$\mathrm{\mu eW^{-1/3}cm^{2/3}}$] & $\gamma$ \\ 
\hline 
		$W_\mathrm{R}$& 78 & 1689&$1627\pm7$ &   $10.9\pm0.7$& $-$  &$0.85^{\pm0.01}$ \\ 
		$W_\Gamma^\mathrm{QD}$& 51& 1725&$1651\pm5$ & $11.5\pm0.5$ & $-$& $0.91^{\pm0.02}$ \\ 
		$W_L^\mathrm{QD}$ & 36& 1755&$1695\pm3$ & $9.3\pm0.4$ & $-$&  $1.23^{\pm0.05}/0.94^{\pm0.03}$ \\ 
		$W_X^\mathrm{IL}$ & 25& 1784& $1729\pm2$ & $8.5\pm0.3$ & $-$&  $2.1^{\pm0.2}$/$0.66^{\pm0.02}$ \\ 
		$W_{\Gamma^*}^\mathrm{QD}$ & 24&1838&$1767\pm1$ & $5.8\pm0.1$ & $-$&  $1.06^{\pm0.08}$ \\ 
		\hline
		$C_\mathrm{R}$ & 67&1645& $1570\pm4$ & $11.1\pm0.3$ & $-$&  $0.85^{\pm0.01}$ \\ 
		$C_L^\mathrm{QD}$& 57 &1701& $1652\pm3$ ($1668\pm3$)& $6.9\pm0.3$ ($4.5\pm0.4$)& ($0.22\pm0.04$)&  $0.60^{\pm0.01}$ \\ 
		$C_\Gamma^\mathrm{QD}$ & 35& 1732&$1700\pm1$ ($1706\pm2$ )& $4.7\pm0.1$ ($3.7\pm0.3$)& ($0.09\pm0.02$)&  $0.61^{\pm0.02}$ \\ 
		$C_X^\mathrm{IL}$& 26 &  1772&$1735\pm1$ ($1743\pm4$ ) & $4.0\pm0.7$ ($3.0\pm0.4$)& ($0.08\pm0.03$)&  $2.4^{\pm0.1}/0.93^{\pm0.03}$ 
\end{tabular} 
\end{ruledtabular}}
\hfill{}
\end{table*}
\end{center}
On the other hand, the energy blueshift with excitation and activation energies are clearly connected to QDs states. This is true as well for sample $S_\mathrm{cap}$ and its emission bands $C_\Gamma^\mathrm{QD}$ and $C_{L}^\mathrm{QD}$. By evaluation of integrated PL intensity in the present excitation density range, we found two regimes,~i.e., below (A) and above (B) $D=1\,\mathrm{W/cm}^2$. We use the similarity in the position of the boundary between these regimes to recognize the character of transitions. Based on similar values of $\gamma$ for $W_X^\mathrm{IL}$ with transition $O_X^\mathrm{IL}$ and $W_L^\mathrm{QD}$ with $O_\mathrm{1R}^\mathrm{IL}$ of sample $S_\mathrm{w/o}$ we identified these bands as ZPL and (partially) phonon-assisted recombination of strained $X_{xy}$ electrons to $\Gamma$-heavy holes in GaAs IL, which was confirmed by comparison with energies extracted from band-schemes in Fig.~\ref{pic:Swith_int}, see Fig.~\ref{fig:kp_results}~(a). It is important to point out that the transition energies are reduced compared to $S_\mathrm{w/o}$, due to the relaxation of the strain in IL (strain is reduced from $-3.4\%$ to $-2.7\%$, see Tab.~\ref{tab:Strain}) due to presence of QDs, and also via exciton localization arising from material intermixing
(represented by the Urbach energy $U_\mathrm{e}+U_\mathrm{h}=9$\,meV which is an order of magnitude larger than that for ${S_\mathrm{w/o}}$). The integrated PL intensity was fairly well fitted by the linear function in the whole range and the obtained value of $\gamma\sim 1$ corresponds to exciton transitions, see Fig.~\ref{pic:Swith_int}~(c).

The comparison of emission energies obtained for $D=0.1~\mathrm{W/cm^2}$ with our calculations for various Ga contents in In$_{1-x}$Ga$_x$As QDs in Fig.~\ref{fig:kp_results}~(a) show a reasonably good agreement for an In$_{0.44}$Ga$_{0.56}$As QD. $W_\Gamma^\mathrm{QD}$ is identified as a transition between $\Gamma$-electrons and holes inside the QD. The large value of FWHM of 78~meV in comparison to the energy scale of the whole spectrum averts the determination of the nature of $W_\mathrm{R}$, which could be associated with phonon-replicas of the $\Gamma$ exciton or transitions between QDs-GaP interface electrons and holes localized in the QD. However, this broad band can be also affected by impurities and indirect transitions in GaP. The width of this band might be also associated to small fluctuations in the size and material composition of the QDs in the measured ensemble. But we tend to identify it as a phonon-replica of $W_X^\mathrm{IL}$, based on comparison with $S_\mathrm{w/o}$. Because $W_{\Gamma^*}^\mathrm{QD}$ appears in our spectra only after the blueshift of other bands start to saturate, we assume its origin is that of a higher excited multi-particle complex,~e.g., charged exciton or excited state of the $\Gamma$-exciton, and we, hence, label it as $\Gamma^*$. 

For all bands along within the whole PL spectrum, a blueshift of the emission $\Delta E$ with increasing pumping of more than 48~meV was observed, saturating for $D>10$\,$\mathrm{W/cm^2}$. The observed blueshift is commonly regarded as a sign of type-II (spatially indirect) transitions. For QDs homogeneously surrounded by substrate material, such type-II transitions can be reasonably well described by $\Delta E \propto D^{1/3}$~\cite{Ledentsov1995,Kuokstis2002,Jo2012,alonso_optical_2007}. This analysis, however, fails in case of a QD positioned on an IL of different material, and more elaborate models are needed. We have recently proposed one in Ref.~\cite{Klenovsky2017}, based on a semi-self-consistent configuration interaction method (SSCCI) with an additional charge background due to impurities. In this model, the blueshift is then a result of ``squeezing" of the wavefunction outside of the dot towards that inside. While such a ``squeezing" and related energy shift is very well known for type-II Sb-systems grown on GaAs substrate~\cite{Gradkowski2012,llorens_type_2015}, they occur also for type-I recombination where the hole wavefunction is partially extending from the QD towards IL, which is the case of our samples (see supplementary, where also results for one calculation of type-I QD using SSCCI model with parameters from~\cite{Kuokstis2002,landoltbornstein} is presented). For convenience, we adopt here for the analysis of the energy shift as a result of excitation density dependence the analytical model of Abramkin \textit{et al.}~\cite{Abramkin_blueshift_analytical} also for QD samples.

The analysis in Fig.~\ref{pic:Swith_int} shows that the blueshift conforms to being due to trap states with Urbach energies around 10~meV and the band-bending, characterized by the $\beta$ parameter, is negligible. Therefore we assign all observed bands to be based on type-I confinement which is in agreement with the type of band-alignment in our simulations. From the temperature analysis of energy shifts of $W_\Gamma^\mathrm{QD}$, $W_L^\mathrm{QD}$, and $W_X^\mathrm{IL}$ bands, we can observe that the slopes of energy change with temperature described by parameters $\alpha$ is similar to bulk values of GaAs or to a combination of that with InAs pointing to the contribution of both GaAs IL and (InGa)(AsSb) QDs. The decrease of $\beta_\mathrm{V}$ with respect to bulk values listed in Tab.~\ref{tab:QDs_temperature} is probably related to quantum confinement. Arrhenius plots, similarly to the case of sample $S_\mathrm{w/o}$, points to low-temperature quenching via impurities with activation energies around 10$\,$meV, close to the deduced localization energy $\sigma_\mathrm{E}\approx 5\,$meV, as discussed before, and activation energies due to electron escape from $X_{xy}$ state from IL to bulk. Moreover, escape energies of 77 and 147$\,$meV were determined which are close to escape energies from QD to IL (from $\Gamma$ 174$\,$meV, from $L$ 154$\,$meV) or to bulk ($L$ 62$\,$meV).

\begin{figure}[!ht]
	\centering
	\includegraphics[width=1\linewidth]{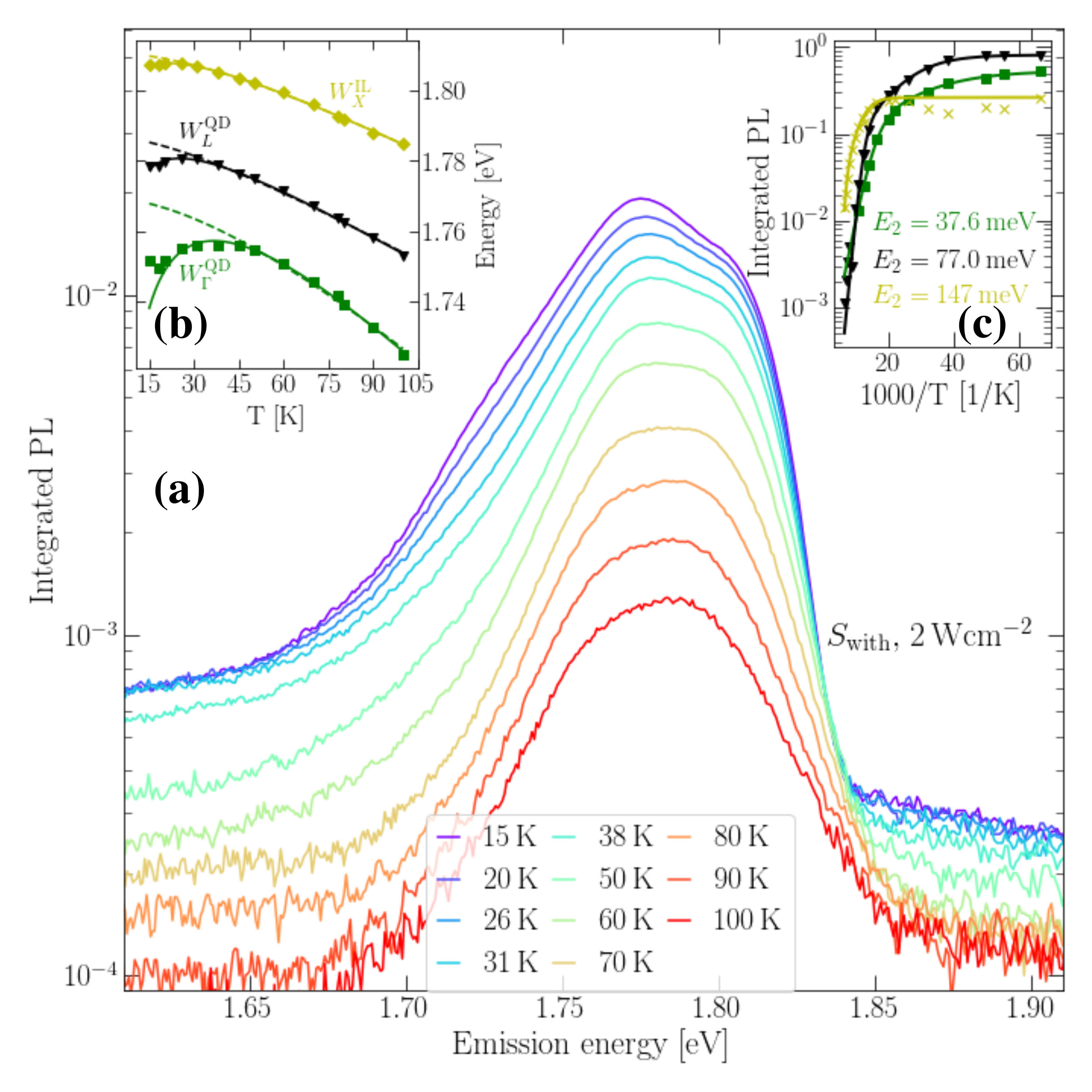}
	\caption{$S_\mathrm{with}$: (a) measured PL at $D=2\,\mathrm{Wcm^{-2}}$ for temperatures between 15\,K and 100\,K. In inset (b) we show the emission energy of $W_\Gamma^\mathrm{QD}$--$W_X^\mathrm{IL}$ (symbols) as a function of temperature fitted by the Varshni Eq.~(\ref{eq:Varshni}), solid curve, [modified Varshni Eq.~(\ref{eq:Varshni-like}), dashed curve] model. In inset (c), integrated PL of individual transitions $W_\Gamma^\mathrm{QD}$--$W_X^\mathrm{IL}$ (symbols) fitted (solid curves) by the Boltzmann model, Eq.~(\ref{eq:Arrhenius}) is shown along with the high temperature activation energies $E_2$.}\label{pic:Swith_temp}
\end{figure}

\begin{center}
\begin{table*}[ht]
{\small
\hfill{}
\caption{Parameters from temperature evolution of emission energies of samples $S_\mathrm{with}$ and $S_\mathrm{cap}$ analysed with the Varshni model (the Varshni parameters $E_{V,0}$, $\alpha$ and $\beta_\mathrm{V}$) with Eliseev thermalization correction $\sigma_\mathrm{E}$, Eq.~(\ref{eq:Varshni-like}), or without that, by Eq.~(\ref{eq:Varshni}) (brackets) and from evolution of integrated intensity analysed using the Boltzmann model~Eq.~(\ref{eq:Arrhenius}) with two activation processes with activation energies $E_1$, $E_2$ and corresponding ratio of radiative and nonradiative lifetimes $\tau_0/\tau_1^\mathrm{NR}$, $\tau_0/\tau_2^\mathrm{NR}$. For comparison, bulk values taken from Ref.~\cite{Vurgaftman2001} are added. The accuracy of the fitted parameters is better than 3\,\% except values marked by $^*$ which have accuracy $\approx5$\,\%.}
\label{tab:QDs_temperature}
\begin{ruledtabular}
\begin{tabular}{lcccc|cccc}
				transition & $E_{V,0}$ [meV]& $\alpha$ [$10^{-4}\,\mathrm{eVK^{-1}}$]& $\beta_\mathrm{V}$ [K]& $\sigma_\mathrm{E}$ [meV]&  $\tau_0/\tau_1^\mathrm{NR}$ & $E_1$ [meV]& $\tau_0/\tau_2^\mathrm{NR}$ $\times 10^3$ & $E_2$ [meV]\\
		\hline
		$W_\Gamma^\mathrm{QD}$& $1798$ & $4.54$& $11.7$&$4.6$&  $9.1$& $6.9 $& $4.078$& $37.6$\\
		$W_L^\mathrm{QD}$& $1796$ $(1787)$& $4.859$ $(4.876)$& $22.1$ $(48.7)$& $4.50$&  $27.75 $& $11.3 $& $610.0$& $77.0$\\
		$W_X^\mathrm{IL}$ & $1819$ $(1812)$& $3.703$ ($3.874)$& $12.0$ $(45.2)$& $3.57$& $99.0 $& $34.5 $& $1028.5$& $146.73$\\ 
		\hline
		$C_L^\mathrm{QD}$& $1764$ $(1749)$ & $6.27$ $(6.5)$ & $10.0$ $(76.8)$&  $4.2$ &  $38.4$& $11.5 $& $52.4$& $77.0$\\
		$C_\Gamma^\mathrm{QD}$& $1777$ $(1771)$ & $4.225$ $(4.463)$ & $10.4$ $(34.3)$&  $2.03$&  $127$& $19.5 $& $423$& $81.9$\\ 
		$C_X^\mathrm{IL}$ & $1791$ $( 1789)$ & $2.4517$ $ (2.4513)$ & $20.00$ $(27.85)$& $1.36$&  $2.4 $& $5.8 $& $0.043^*$& $244.5$ \\
		\hline
		GaAs, $\Gamma$ ($L$) [$X$]& 1519 (1815) [1981]& 5.405 (6.05) [4.60]& 204& \\
		GaSb, $\Gamma$ ($L$) [$X$]& 812 (875) [1141]& 4.17 (5.97) [4.75]& 140 [94]& \\
		InAs, $\Gamma$ ($L$)& 417 (1133)& 2.76& 93& \\
		InSb, $\Gamma$ ($L$) [$X$]& 235 (930) [630]& 3.20& 170&\\
		GaP, $\Gamma$ ($L$) [$X$]& 2886 (2720) [2350]& 5.77& 372& 

\end{tabular} 
\end{ruledtabular}}
\hfill{}
\end{table*}
\end{center}

\subsection{Sample with capped QDs $\mathbf{S_\mathrm{cap}}$}
PL spectra of the sample ${S_\mathrm{cap}}$ as a function of both excitation density and temperature were fitted by the sum of 4~Gaussian profiles labeled from smaller to larger mean energy as $C_\mathrm{R}$ to $C_X^\mathrm{IL}$, according to the labeling in Fig.~\ref{fig:PL_int_QDs}~(b), where fits for two different excitation densities are shown. Emission energy, FWHM, and PL intensity have been investigated, and the character of bands was determined similarly as in the case of sample ${S_\mathrm{with}}$. Comparison of the emission energies for $D=0.1~\mathrm{W/cm^2}$ with $\mathbf{k \cdot p}$ calculations in Fig.~\ref{fig:kp_results} indicates that $C_\Gamma^\mathrm{QD}$ and $C_L^\mathrm{QD}$ are most probably bands with contribution of $\Gamma$- and $L$-electron-hole transitions in In$_{0.2}$Ga$_{0.8}$As$_{0.84}$Sb$_{0.16}$ QDs. Note that the aforementioned composition of the QDs has been found by matching the experimental results to corresponding ones obtained using {\bf k$\cdot$p} theory. The origin of $C_X^\mathrm{IL}$ as a transition in GaAs IL from $X_{xy}$-electrons to heavy holes has been deduced from the ${\bf k\cdot p}$ band-scheme. It is important to point out that the energy of the $X_{xy}$-hole in Fig.~\ref{fig:kp_results} is underestimated because we used the energy values of the band-edges in GaAs IL, and not those of the confined states. This assignment is supported by the similarity of the integrated PL intensity with excitation energy where, for that band, three segments have been observed (also similar to the case of $S_\mathrm{w/o}$). Unfortunately, it was not possible to investigate the $C_\mathrm{R}$ band in more detail owing to its width of 67\,meV for $D=0.1\,\mathrm{W/cm^2}$ and overall lower emission intensity. We assume that this band arises from the recombination of excitons from QD regions with a slightly varying material concentration towards the capping layer, mixed with phonon-assisted transitions of QDs and IL states convoluted with DAP emission.

\begin{figure*}[!]
	\centering
	\includegraphics[width=1.0\linewidth]{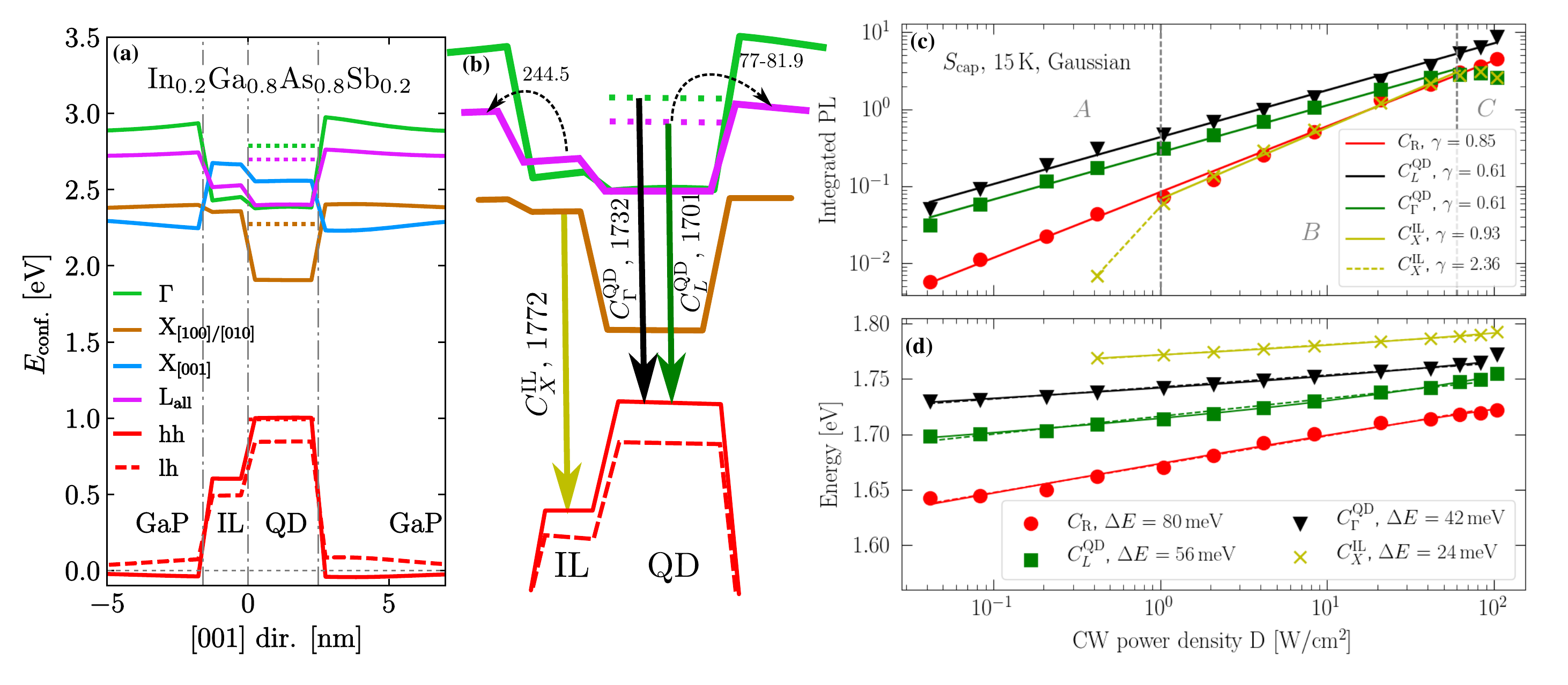}
	\caption{$S_\mathrm{cap}$: (a) Calculated bandscheme of a $\mathrm{In_{0.2}Ga_{0.8}As_{0.8}Sb_{0.2}}$ QD using nextnano++~\cite{Birner:07} simulation suite with single particle eigenenergies indicated by dotted lines. (b) Detail of the bandscheme with indicated emission recombination paths (solid arrows with transition energies in meV) and escape energies (values in meV) found from temperature dependent PL. (c) Integrated PL intensity of individual transitions (symbols) fitted by linear dispersion in log-log scale, corresponding slopes for segments are given in the inset. (d) Emission energies of individual transitions (symbols) vs. excitation density $D$. Values of absolute blueshifts are given in the inset. Solid (dashed) lines represent fits by Eq.~(\ref{eq:PL_intmodel}) with $\beta=0$ ($\beta\neq0$).}    
	\label{pic:Scap_int}
\end{figure*}

For the $C_\mathrm{R}$ band, an energy shift $\Delta E$ of more than 33\,meV is observed (such value has been obtained from energy extrapolation towards the smallest excitation density of $C_X^\mathrm{IL}$). Comparing to $S_\mathrm{with}$, a shift towards larger emission energies with increasing excitation density can be observed, but without saturation above 10\,$\mathrm{W/cm^2}$. Such shift can be described by Eq.~(\ref{eq:PL_intmodel}) by using a bending parameter smaller than 0.3\,$\mu\mathrm{eV}$, which is insignificant in comparison to that in type-II QW systems ($\beta=14\,\mu \mathrm{e W^{-1/3}cm^{2/3}}$ for GaAs/AlAs; $\beta=12\,\mu \mathrm{e W^{-1/3}cm^{2/3}}$ for AlSb/AlAs)~\cite{Abramkin_blueshift_analytical}. We point out that $\beta$ for QD systems is usually not determined, because the type of band-alignment is determined from the magnitude of $\Delta E$ only~\cite{HATAMI1995, Hatami1998}. These negligibly small values of $\beta$ point to the dominant effect of a background potential on the energy shift due to trap states, the so-called ``state-filling effect"~\cite{Abramkin_blueshift_analytical}, and the band-alignment type is found to be type-I. This shift was evaluated by Eq.~(\ref{eq:PL_intmodel}) with $\beta=0$ and Urbach energies of 4--7\,meV. This assignment fully agrees with the theoretical prediction in~\cite{Klenovsky2018_TUB} where the formation of type-II interfaces for QDs to a concentration of Sb around 20\,\% on GaP in contrast with GaAs substrate is unlikely. 

\begin{figure}[!h]
	\centering
	\includegraphics[width=1\linewidth]{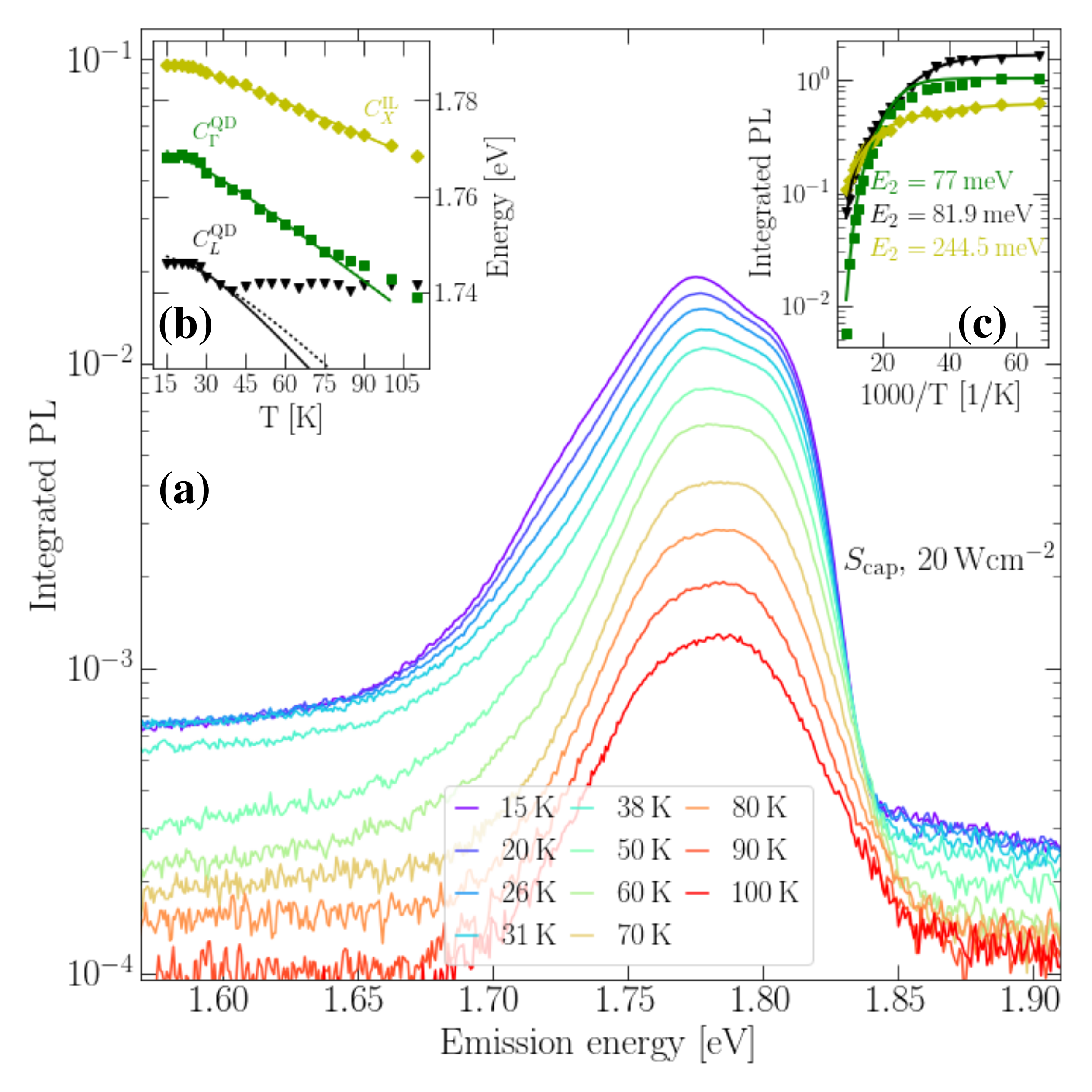}
	\caption{$S_\mathrm{cap}$: (a) PL spectra for $D=20\,\mathrm{Wcm^{-2}}$ for temperatures between 15\,K and 100\,K. The results in panels (b) and (c) are given in the same way as in Fig.~\ref{pic:Swith_temp}.}\label{pic:Scap_temp}
\end{figure}

The radiative recombination of bands $C_L^\mathrm{QD}$--$C_X^\mathrm{IL}$ has been studied as a function of temperature from 15\,K up to 100\,K, as for the sample ${S_\mathrm{with}}$, see Fig.~\ref{pic:Scap_temp}.
First, we observe a thermalization and $\sigma_\mathrm{E}$ is found to be 1--4\,meV. The thermalization process occurs most probably through impurities. Furthermore, the bands are Varsni-like shifted with the rate parameter $\alpha$ close to bulk values. The parameter $\alpha$ for $C_L^\mathrm{QD}$ of $6.5 \cdot10^{-4}\,\mathrm{eVK^{-1}}$ supports our previous assignment,~i.e., that $C_L^\mathrm{QD}$ transition involves states being momentum-indirect and that it originates from a structure with a low amount of In, which pushes $\alpha$ towards smaller values. The $\Gamma-$character of $C_\Gamma^\mathrm{QD}$ is supported by the corresponding value of $\alpha=4.46 \cdot10^{-4}\,\mathrm{eVK^{-1}}$ which is very close to $\Gamma$-Bloch wave bulk parameters.

By comparing the $\alpha$ parameters extracted from PL spectra of sample ${S_\mathrm{cap}}$ with those of ${S_\mathrm{with}}$, we observe that $\alpha$ for transition involving $\Gamma$-electrons in QD is slightly reduced for ${S_\mathrm{cap}}$ while $\alpha$ for $L$ is increased. This can be quantitatively understood as an effect of an increasing amount of Sb in combination with a decrease of In in the QDs due to the Sb-P-As exchange processes (discussed in TEM results in Sec.~\hyperref[sec:Fabrication]{Sample fabrication and structural characterization}). A similar explanation applies for the decreasing $\alpha$ of $X_{xy}$ for $S_\mathrm{cap}$.

The radiative recombination is quenched due to impurities (activation energies $E_1\sim10$\,meV), similarly to the previous samples. Using Arrhenius plots, in Fig.~\ref{pic:Scap_temp}~(c) we also extracted the energies of $\Gamma$-electrons confined in the QDs ($E_2=77$ or $82$\,meV from $C_L^\mathrm{QD}$ and $C_\Gamma^\mathrm{QD}$ respectively, the theoretical value from Fig.~\ref{pic:Scap_int} is 85\,meV) and the escape energy of electron from $L$ to GaP substrate ($E_2=245$\,meV, the theoretical value from Fig.~\ref{pic:Scap_int} is 222\,meV).
The small discrepancy could arise because we compare experiment with the theoretical values taken from QDs with slightly different composition, In$_{0.2}$Ga$_{0.8}$As$_{0.9}$Sb$_{0.1}$, which is the closest match.

\section{Polarization of emission}

\begin{figure*}[!]
	\centering
	\includegraphics[width=1\linewidth]{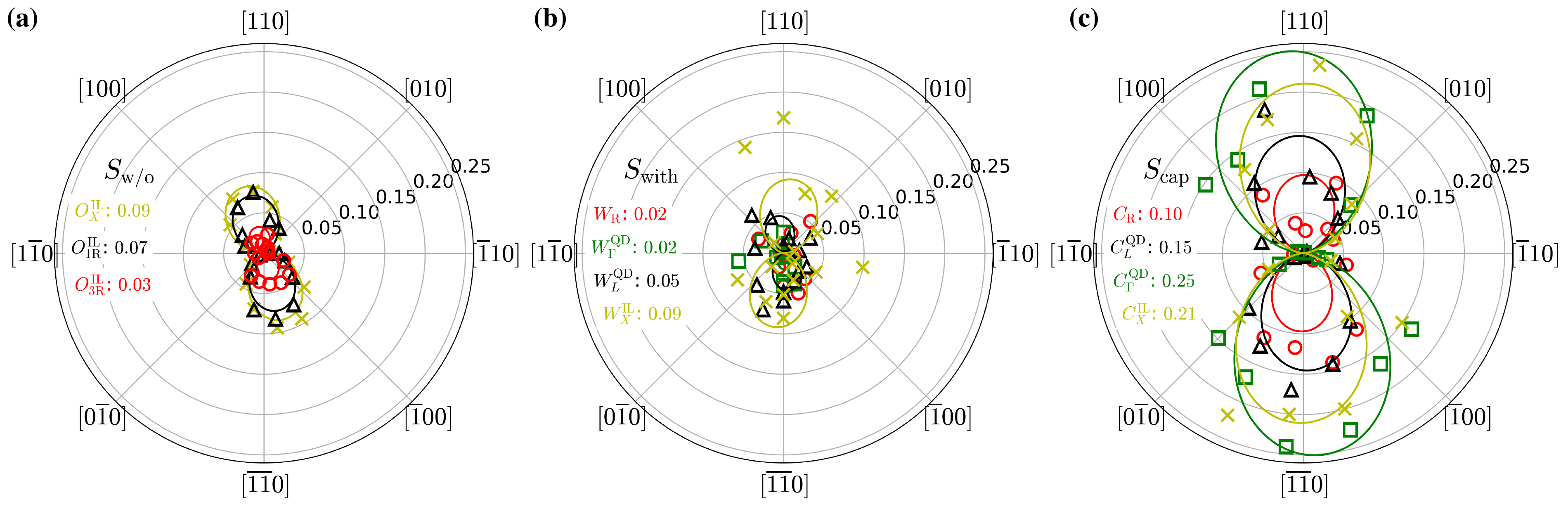}
	\caption{From left to right we show the polar graphs of $C(\theta)$ for samples {\bf{a)}} ${S_\mathrm{w/o}}$,  {\bf{b)}} ${S_\mathrm{with}}$, and  {\bf{c)}} $S_\mathrm{cap}$, respectively. Individual bands of PL spectra for each sample are represented by different symbols, consistently with labelling in previous figures.}\label{pic:Polarization}
\end{figure*}

In our experiments, both the excitation beam and the detected PL radiation propagate perpendicularly to the sample surface, and we analyze the latter by a rotating half-wave plate followed by a~fixed linear polarizer. The angle between the crystallographic direction~[110] and the polarization vector is denoted $\theta$ in the following. 

The emission of strained GaAs layer on GaP was previously studied in connection with quantum efficiency and spin-polarization as a function of strain~\cite{MAIR_1996} or thickness~\cite{Maruyama_3nmGaAsonGaP_POL}. Predominantly anisotropy of the emission along the two orthogonal directions [110] and [1-10] of only few percent was found to be caused by an asymmetric strain relaxation~\cite{MAIR_1996} for emission energy around 1.8\,eV~\cite{Maruyama_3nmGaAsonGaP_POL}. Since similar small degrees of polarization of the emitted light were expected also for our samples, we visualize our results in terms of the degree of polarization~\cite{Klenovsky2015}
\begin{equation}
C(\theta)=\frac{I(\theta)-I_\mathrm{min}}{I_\mathrm{max}+I_\mathrm{min}},
\end{equation}
where $I_\mathrm{min}$ and $I_\mathrm{max}$ are extreme values of PL intensity $I(\theta)$; $\theta$ denotes the angle. Note, that for angle $\theta_\mathrm{max}$, such that $I(\theta_\mathrm{max})=I_\mathrm{max}$, the previous relation gives the maximum degree of polarization $C(\theta_\mathrm{max})=C_\mathrm{max}$ (values in the polar graphs in Fig.~\ref{pic:Polarization}).

The emission radiation from samples ${S_\mathrm{with}}$ and ${S_\mathrm{cap}}$ is polarized along the [110]~crystallographic direction, in agreement with results on type-I InAs/GaAs QDs~\citep{HumPhysE} where the polarization anisotropy of $I(\theta)$ is given predominantly by the orientation of the wavefunction of hole states. Based on that, and noting the results of Ref.~\citep{Klenovsky2015}, we conclude that the transitions in the studied samples agree with a type-I band alignment.

The sample ${S_\mathrm{with}}$ has $C_\mathrm{max}$ around 0.05, which is comparable to that for InAs/GaAs QDs, where single-particle wavefunctions are located approximately in similar locations around the QD and also to the results of Ref.~\cite{Klenovsky2018_TUB}. On the other hand, antimony from the GaSb capping in sample ${S_\mathrm{cap}}$ positions the wavefunctions of electrons and holes slightly further apart from each other, therefore, $C_\mathrm{max}$ increases up to almost 0.25. We note that this result, along with the polarization of emission (rotated by 90\,$^\circ$) and the system still displaying in type-I confinement, shows that the presence of a Sb-rich layer above QDs in ${S_\mathrm{cap}}$ causes the hole states to be oriented towards the Sb layer (or even partly leak out there). Such a scenario is then similar to InAs QDs capped by thin GaAsSb layers, see Ref.~\cite{Klenovsky2015}.
 
Even though we have performed an advanced analysis, we cannot clearly distinguish between QDs and IL bands, due to their energetic overlap and additional phonon induced-thermal broadening. In order to better address the contributions of IL and QDs, we recommend to perform resonant spectroscopy measurements, which allow to distinguish the bands, thanks to different resonant frequencies, while at the same time observing their indirect momentum~\cite{Rauter_indirectQD}, or to use interferometry, which allows to discern the bands having different lifetimes (topic of our current study in~\cite{Steindl2019_TRPL}).
\section{Conclusions}
In conclusion, by combining excitation and temperature resolved PL with $\mathbf{k\cdot p}$ calculations, we have investigated the optical properties of III--V nanostructures grown on GaP (001) substrates, in relation to their electronic structure. We have compared the results for three different structures grown by MOCVD, which differed in sequential adding of GaAs IL, (InGa)(AsSb) QDs and thin GaSb capping layer. For these systems, type-I (spatially direct) momentum-direct and indirect transitions have been found, with significant blueshift with increasing pumping for samples containing QDs. HRTEM measurements have been performed in order to determine the QD morphology and stoichiometry and used as a basis for the theoretical simulations. The growth of QDs leads to modification of the hydrostatic strain in the GaAs IL, confirmed by a combination of Raman and photoreflectance spectroscopy. For a strained GaAs layer only, the dominant emission is originating from $X_{xy}$-$\Gamma$ electron-hole transition (1.855\,eV), and is assisted by GaAs and GaP phonon-replicas.
The subsequent (InGa)(AsSb) QDs growth leads to a better strain relaxation, revealed by a slight redshift of the transitions involving $X_{xy}$-electrons in the layer. Additionally, the QD growth introduces transitions involving $\Gamma$-electrons (1.725\,eV) and $L$ (1.755\,eV), which are both strongly blueshifted (more than 50\,meV) with increasing excitation density. Such blueshift can be attributed to defects in the QD region, which are created during strain relaxation. Moreover, saturation has been observed starting from excitation densities of more than 10\,$\mathrm{W/cm^2}$. Overgrowing the QDs by a GaSb cap layer results in a material intermixing via Sb-As exchange reactions. Such a process effectively modifies the overall composition of QDs. This leads to an energy swapping of $\Gamma$ (1.732\,eV) and $L$ (1.701\,eV) bands, and to an enhanced leakage of the electron wavefunctions out of the QD body - also confirmed by analyzing the polarization of the emission - and leading to a twice weaker emission intensity. 
We observe a large blueshift with increasing excitation density for the QDs studied here, even though they present a type-I band-alignment. We find that addressing the phenomenon of blueshift with increasing optical pumping to a type-II band alignment is insufficient. Therefore, a more consistent approach based on the comparison of the evolution of the shift with excitation has been employed, which followed the analytical model given by Abramkin \textit{et al.}~\cite{Abramkin_blueshift_analytical}. That allows to distinguish between band-bending and state filling which occurs due to impurities, or more generally to employ a numerical model based on the self-consistent cycle, such as a semi-self-consistent configuration interaction method~\cite{Klenovsky2017}.





\section{Acknowledgements}
The authors would like to acknowledge Jan Michali\v{c}ka for the TEM measurements, J\"urgen Christen for fruitful discussion about spectral shape-line models and Gaspar Armelles for providing insight in the early GaAs/GaP optical studies.
P.S. is Brno Ph.D. Talent Scholarship Holder--Funded by the Brno City Municipality. 
E.M.S. thanks the DFG (Contract No. BI284/29-2).
A part of the work was carried out under the project CEITEC 2020 (LQ1601) with financial support from the Ministry of Education, Youth and Sports of the Czech Republic under the National Sustainability Programme II. This project has received national funding from the MEYS and funding from European Union's Horizon 2020 (2014-2020) research and innovation framework programme under grant agreement No 731473. The work reported in this paper was (partially) funded by project EMPIR 17FUN06 Siqust. This project has received funding from the EMPIR programme co-financed by the Participating States and from the European Union’s Horizon 2020 research and innovation programme. The work was also partially founded by Spanish MINEICO EUIN2017-88844.


\bibliography{paper_TUB_PL_QDs.bbl}

\begin{thebibliography}{128}
\bibitem [{Not({\natexlab{b}})}]{NotePCM}%
  \BibitemOpen
  \href@noop {} {\enquote {\bibinfo {title} {{Considering the QD high density
  and rather of homogeneous size together with collecting conventional Raman
  spectra at room temperature we approximate them by a thin layer of a
  thickness of 2.5\,nm.}}}\ } ({\natexlab{b}})\BibitemShut {NoStop}%
\end{thebibliography}

\end{document}